\documentclass[aps,prd,11pt,notitlepage,longbibliography,nofootinbib,tightenlines,preprintnumbers]{revtex4-1}
\pdfoutput=1
\usepackage{amsmath,amssymb,amsfonts}
\usepackage{graphicx}
\usepackage{color}
\usepackage{tikz}
\usepackage{dsfont}
\usepackage[pdftex]{hyperref}
\hypersetup{colorlinks=true, linkcolor=darkred, citecolor=blue, linktoc=page}
\definecolor{darkred}{rgb}{0.8,0.1,0.1}

\usepackage{diagbox}

\usepackage{subfigure}

\def\RR{{\mathds{R}}}

\makeatletter\def\l@subsubsection#1#2{}%
\makeatother

\newcommand{\nocontentsline}[3]{}
\newcommand{\tocless}[2]{\bgroup\let\addcontentsline=\nocontentsline#1{#2}\egroup}

\def\Im{\mathop{\rm Im}}
\def\Re{\mathop{\rm Re}}

\begin{document}

\title{Islands and Page curves in 4d from Type IIB}

\author{Christoph F.~Uhlemann} 
\email{uhlemann@umich.edu}

\affiliation{Leinweber Center for Theoretical Physics, Department of Physics	\\
	University of Michigan, Ann Arbor, MI 48109-1040, USA}

\preprint{LCTP-21-09}

\begin{abstract}
Variants of the black hole information paradox are studied in Type IIB string theory setups that realize four-dimensional gravity coupled to a bath. The setups are string theory versions of doubly-holographic Karch/Randall brane worlds, with black holes coupled to non-gravitating and gravitating baths. The 10d versions are based on fully backreacted solutions for configurations of D3, D5 and NS5 branes, and admit dual descriptions as $\mathcal N=4$ SYM on a half space and 3d $T_\rho^\sigma[SU(N)]$ SCFTs. Island contributions to the entanglement entropy of black hole radiation systems are identified through Ryu/Takayanagi surfaces and lead to Page curves. Analogs of the critical angles found in the Karch/Randall models are identified in 10d, as critical parameters in the brane configurations and dual field theories.
\end{abstract}

\maketitle

\bigskip
\tableofcontents
\bigskip

\setlength{\parskip}{1.6pt}

\section{Introduction and Summary}

A recent advance in the understanding of black holes are the computations \cite{Penington:2019npb,Almheiri:2019psf} of the time evolution of the entanglement entropy between a holographic black hole system and an external bath to which the black hole is coupled.
A crucial ingredient in these computations are entanglement islands -- contributions to the entanglement entropy (EE) from regions that are disconnected and can be far away from the bath \cite{Almheiri:2019hni,Almheiri:2019yqk,Almheiri:2019psy,Almheiri:2019qdq,Penington:2019kki}.
These contributions become dominant at late times and lead to Page curves for the time evolution of the entropy, in line with expectations based on unitarity. Reviews can be found in \cite{Almheiri:2020cfm,Raju:2020smc}.

The discussions so far are largely based on bottom-up models and on low-dimensional theories where the features of gravity are qualitatively different. 
A prominent role is played by Karch/Randall models \cite{Karch:2000ct,Karch:2000gx}.
The special case of a Karch/Randall model with a tensionless end-of-the-world brane, discussed in \cite{Geng:2020qvw}, can be embedded into Type IIB string theory as an orbifold of $AdS_5\times S^5$. But that case is somewhat peculiar in that the 4d graviton has a mass that can not be separated from the UV cut-off in the 4d gravitational description.

The aim of the present work is to demonstrate in a UV-complete string theory setting the emergence of entanglement islands and Page curves for black holes in four-dimensional theories of gravity in which the graviton mass can be controlled, including theories with massless gravitons.
Starting point are the discussions of islands and Page curves in general 5d Karch/Randall models \cite{Chen:2020uac,Chen:2020hmv,Geng:2020qvw,Geng:2020fxl,Rozali:2019day}, which can be used to model gravitating systems coupled to non-gravitating and gravitating baths.
These models have the appealing feature that the quantum extremal surfaces \cite{Engelhardt:2014gca,Faulkner:2013ana} exhibiting island contributions are entirely geometrized, due to the doubly-holographic nature of these models. This allows for the identification of entanglement islands through classical Ryu/Takayanagi surfaces \cite{Ryu:2006bv}. 
We will uplift the discussions in these bottom-up models to Type IIB string theory, to provide UV completions and concrete holographically dual QFTs.

The string theory constructions are based on holographic duals for 4d boundary CFTs and for 3d SCFTs  engineered by configurations of D3, D5 and NS5 branes \cite{Gaiotto:2008sa,Gaiotto:2008sd,Gaiotto:2008ak}. Holographic duals for large classes of such theories were constructed in \cite{DHoker:2007zhm,DHoker:2007hhe,Aharony:2011yc,Assel:2011xz}, and they provide natural string theory realizations of the Karch/Randall models with non-gravitating and gravitating baths.
We will study quantum extremal/minimal surfaces in these solutions and identify quantities that exhibit Page curve behavior.
The key findings of \cite{Geng:2020fxl,Geng:2020qvw}, such as the existence of critical brane angles separating different phases of minimal surfaces, will find string theory realizations. We will also identify 10d versions of the ``left/right EE'' that was found to exhibit Page curve behavior in the 5d models with gravitating bath, where the usual notion of geometric EE becomes subtle.

In the following we will first review relevant aspects of the discussion in the Karch/Randall models to set the stage and then summarize the main results of this paper.

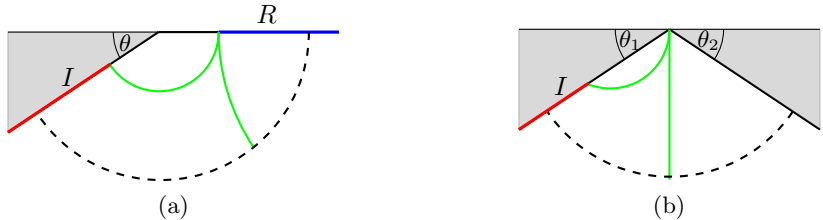
\begin{figure}
	\subfigure[][]{\label{fig:KR-nongrav}
		\begin{tikzpicture}[scale=0.8]
			\draw (-2.5,0) -- (0,0);
			\draw[thick](0,0) -- (1,0);
			\draw[very thick,blue] (1,0) -- (3,0);
			
			\node [anchor=south] at (1.8,0) {\small $R$};
			
			\draw[thick] (0,0) -- (-2.5,-2/3*2.5);
			
			\draw[thick,green] (1,0) arc (180:214:98pt);
			\draw[thick,green] (1,0) arc (0:-148:28pt);
			\draw[thick,dashed,black] (2.5,0) arc (0:-146:70pt);
			
			\draw [fill=gray,opacity=0.3] (0,0) -- (-2.5,0) -- (-2.5,-2/3*2.5)--(0,0);
			
			\node at (-1.5,-0.75) {\small $I$};
			\draw[very thick,red] (-0.8,-2/3*0.8) -- (-2.5,-2/3*2.5);
			
			\node at (-0.55,-0.18) {\footnotesize $\theta$};
			\draw (-0.75,0) arc (180:210:25pt);
		\end{tikzpicture}
	}
	\hskip 20mm
	\subfigure[][]{\label{fig:KR-grav}
		\begin{tikzpicture}[scale=0.8]
			\draw (-2.5,0) -- (2.5,0);

			\draw[thick] (0,0) -- (-2.5,-2/3*2.5);
			\draw[thick] (0,0) -- (2.5,-2/3*2.5);
			
			\draw[thick,green] (0,0) -- (0,-2.5);
			\draw[thick,green] (0,0) arc (0:-113:28pt);
			\draw[thick,dashed,black] (2.05,-2/3*2.05) arc (-34:-146:70pt);
			
			\draw [fill=gray,opacity=0.3] (0,0) -- (-2.5,0) -- (-2.5,-2/3*2.5)--(0,0);
			\draw [fill=gray,opacity=0.3] (0,0) -- (2.5,0) -- (2.5,-2/3*2.5)--(0,0);
			
			\node at (-1.8,-0.95) {\small $I$};
			\draw[very thick,red] (-1.35,-2/3*1.35) -- (-2.5,-2/3*2.5);
			
			\node at (-0.65,-0.18) {\footnotesize $\theta_{1}$};
			\node at (0.65,-0.18) {\footnotesize $\theta_2$};
			\draw (-0.9,0) arc (180:215:25pt);
			\draw (0.9,0) arc (0:-35:25pt);
		\end{tikzpicture}
	}
	
	\caption{
		Left: Karch/Randall model for non-gravitating bath. The figure shows part of $AdS_5$ with the ETW brane cutting off the shaded region.
		The dashed curve is the black hole horizon and $R$ is the radiation region (blue). 
		The green curve ending on the horizon represents the HM surface; 
		the green curve extending from the boundary of $R$ to the ETW brane is the island surface. $I$ is the island (red).
		Right: For a gravitating bath a second ETW brane is introduced, leaving only a 3-dimensional part of the conformal boundary.
		\label{fig:KR}}
\end{figure}

\medskip
\textbf{Islands and Page curves in Karch/Randall models:} 
The Karch/Randall models for 4d gravity coupled to a non-gravitating bath 
are based on a part of  $AdS_5$ cut off by an end-of-the-world (ETW) brane extending along an $AdS_4$ slice (fig.~\ref{fig:KR-nongrav}). 
The conformal boundary is cut off at the point where it is intersected by the ETW brane, so that these models are holographically dual to boundary conformal field theories (BCFTs) (see also \cite{Takayanagi:2011zk,Fujita:2011fp}). 
The advantage of these setups from the entanglement islands perspective is that they have 3 holographically related descriptions:
\begin{itemize}
	\setlength{\parskip}{0 pt}
	\item[(a)] Einstein gravity on (asymptotically) AdS$_5$ + ETW brane
	\item[(b)] a 4d CFT with UV cut-off $+$ gravity on (asymptotically) AdS$_4$, 
	coupled via transparent\\ boundary conditions at the boundary of AdS$_4$ to a 4d CFT on half of $\RR^{1,3}$
	\item[(c)] a non-gravitational 4d CFT on half of $\RR^{1,3}$ coupled to 3d boundary degrees of freedom
\end{itemize}
These descriptions can be understood to arise from applying AdS/CFT twice: description (b) is obtained by converting the 3d boundary degrees of freedom in (c) to a gravitational theory on AdS$_4$, and description (a) geometrizes the entire BCFT.

Description (b) is the one of interest for the black hole information paradox. 
To pose the paradox, the $AdS_4$ slices are replaced by $AdS_4$ black holes. 
This realizes a black hole on the ETW brane and on the remaining half of the conformal boundary of $AdS_5$, which serves as bath. 
It can be interpreted as coupling the gravity system on the ETW brane to a bath at the same temperature as the black hole.
To quantify the entropy of the radiation one picks a region far in the bath system and computes its EE.
One type of surface relevant for computing the EE holographically are Hartman-Maldacena (HM) surfaces \cite{Hartman:2013qma}, which connect the boundary of the radiation region to the corresponding point in the thermofield double.
Due to the stretching of the space behind the horizon the area of these surfaces grows in time,
suggesting an unbounded growth of the entropy. This is the version of the information paradox described in \cite{Almheiri:2019yqk}.
The paradox is resolved by the existence of ``island minimal surfaces'' that stretch from the bath into the gravity system (fig.~\ref{fig:KR-nongrav}). 
The part of the ETW brane near the black hole that is captured by the surface constitutes the island contribution. Its computation is entirely geometrized through the existence of the 5d bulk.
The area of the island surfaces is constant in time, which limits the growth of the entropy and leads to Page curves.
As emphasized in \cite{Geng:2020qvw}, the graviton is generically massive in models with a non-gravitating bath. 

A gravitating bath can be realized by introducing a second ETW brane as bath (fig.~\ref{fig:KR-grav}) \cite{Geng:2020fxl}.
This modifies description (b) to now comprise two CFTs coupled to gravity on distinct $AdS_4$ spaces, and coupled to each other at the conformal boundaries.
Description (c) is reduced to a 3d CFT.
Since both ETW branes have dynamical gravity, a conventional geometric EE can not be defined on the second ETW brane.
If one allows the end points of minimal surfaces on both ETW branes to be chosen dynamically, the surfaces can settle on the horizon and lead to a flat entropy curve, in line with the general arguments of \cite{Laddha:2020kvp}.
The quantity that was found to exhibit Page curve behavior in \cite{Geng:2020fxl} instead corresponds to minimal surfaces anchored at the remaining point of the conformal boundary of $AdS_5$, and was interpreted as EE between defect degrees of freedom represented by the left and right ETW branes.
The form of the entropy curve was found to have interesting dependence on the ETW brane angles, as will be discussed in more detail below.

\medskip
\textbf{Islands and Page curves in Type IIB:}
In this work we will study 10d  string theory versions of the Karch/Randall models
and show that the qualitative features captured by the bottom-up models are realized in a  UV-complete theory of quantum gravity.
We will discuss black holes coupled to non-gravitating and to gravitating baths, realized through 10d black hole solutions based on the $AdS_4\times S^2\times S^2\times\Sigma$ solutions of Type IIB constructed in \cite{DHoker:2007zhm,DHoker:2007hhe,Aharony:2011yc,Assel:2011xz}.

\begin{figure}
	\begin{tikzpicture}
		\shade [ left color=blue! 0, right color=blue! 20] (-2.2,0)  rectangle (0,-2);
		\shade [ right color=blue! 0, left color=blue! 20] (0,0)  rectangle (2.2,-2);
		
		\draw[thick] (-2.2,0) -- (2.2,0);
		\draw[thick] (-2.2,-2) -- (2.2,-2);
		\draw[dashed] (-2.2,0) arc (160:200:85pt);
		\draw[dashed] (2.2,0) arc (20:-20:85pt);
		
		\node at (1.7,-0.6) {$\Sigma$};
		\node at (2.3,-2.3) {\small $x\rightarrow+\infty$};
		\node at (-2.3,-2.3) {\small $x\rightarrow-\infty$};
		\node at (3.0,-0.65) {\small $AdS_5$};
		\node at (3.0,-1) {\small $\times$};
		\node at (3.0,-1.35) {\small $S^5$};
		
		\draw[thick] (0.2,-0.08) -- (0.2,0.08);
		\draw[thick] (-0.2,-0.08) -- (-0.2,0.08);
		\draw[thick] (0,-0.08) -- (0,0.08) node [anchor=south] {\small NS5};
		\draw[thick] (0.2,-2-0.08) -- (0.2,-2+0.08);
		\draw[thick] (-0.2,-2-0.08) -- (-0.2,-2+0.08);
		\draw[thick] (0,-1.92) -- (0,-2.08) node [anchor=north] {\small D5};
		
		\node at (-1,-1.75) {\small $y=0$};
		\node at (-1,-0.25) {\small $y=\frac{\pi}{2}$};
		
	\end{tikzpicture}
\qquad\qquad
	\begin{tikzpicture}[y={(0cm,1cm)}, x={(0.707cm,0.707cm)}, z={(1cm,0cm)}, scale=1.1]
	\draw[white,fill=gray!100] (0,0,0.5) circle (1.5pt);
	\draw[white,fill=gray!100] (0,0,1.5) circle (2pt);
	
	\draw[thick] (0,-0.39,0) -- (0,1,0);
	\draw[thick] (0,-1,0) -- (0,-0.6,0);
	
	\draw[thick] (0,-0.41,1) -- (0,1,1);
	\draw[thick] (0,-1,1) -- (0,-0.61,1);
	
	\draw[thick] (0,-1,2) -- (0,1,2);
	
	\node at (0,0,2.375) {$\cdots$};
	\draw[thick] (0,-1,2.75) -- (0,1,2.75);
	
	\foreach \i in {-0.075,-0.025,0.025,0.075}{ \draw (-1.1,\i,0.5) -- (0.65,\i,0.5);}
	\foreach \i in {-0.075,-0.025,0.025,0.075}{ \draw (0.76,\i,0.5) -- (1.1,\i,0.5);}
	
	\foreach \i in {-0.05,0,0.05}{ \draw (-1.1,\i,1.5) -- (0.65,\i,1.5);}
	\foreach \i in {-0.05,0,0.05}{ \draw (0.76,\i,1.5) -- (1.1,\i,1.5);}
	
	\foreach \i in {-0.025,0,0.025}{ \draw (0,1.4*\i,0) -- (0,1.4*\i,1);}
	\foreach \i in {-0.05,-0.025,0,0.025,0.05}{ \draw (0,1.4*\i,1) -- (0,1.4*\i,2.05);}
	\foreach \i in {-0.045,-0.015,0.015,0.045}{ \draw (0,1.4*\i,2.7) -- (0,1.4*\i,5);}
	
	\node at (-0.18,-0.18,4) {\small D3};
	\node at (1.0,0.2,0.75) {\footnotesize D5};
	\node at (0,-1.25) {NS5};
\end{tikzpicture}
	
\caption{
	Left: Geometry of $AdS_4\times S^2\times S^2\times\Sigma$ solutions with $\Sigma=\lbrace x+iy\in\mathds{C}\vert \,0\leq y\leq \frac{\pi}{2}\rbrace$ for non-gravitating baths. On each boundary component an $S^2$ collapses, so the 10d geometry is closed.
	D5/NS5 brane sources are located on the $y=0$/$y=\frac{\pi}{2}$ boundaries.
	The limit $x\rightarrow -\infty$ is a regular point of the internal space.
	For $x\rightarrow\infty$ the solutions approach locally $AdS_5\times S^5$; this region corresponds to the conformal boundary in fig.~\ref{fig:KR-nongrav}.
	The ETW brane in fig.~\ref{fig:KR-nongrav} can be seen as effective description for the remaining 10d geometry.
	Right: Associated configuration of D5, NS5 and D3 branes, with D3-branes suspended between 5-branes and semi-infinite D3-branes emerging in one direction. The distribution of 5-brane sources in the supergravity solution encodes how many D5/NS5 branes there are and how the D3-branes  end on them.
	\label{fig:AdS4-sol}}
\end{figure}
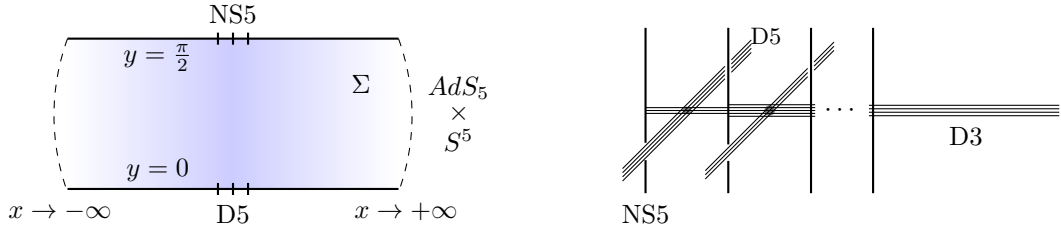

We start the discussion with non-gravitating baths.
The solutions constructed in \cite{DHoker:2007zhm,DHoker:2007hhe,Aharony:2011yc} can be used to describe semi-infinite D3-branes terminating on a system of D5 and NS5 branes with additional D3-branes suspended between the 5-branes.
The brane configurations engineer $\mathcal N=4$ SYM on a half space, corresponding to the semi-infinite D3-branes, coupled to a 3d SCFT on the boundary, corresponding to the D3-branes suspended between the D5 and NS5 branes. 
The structure of the supergravity solutions and brane setups is illustrated in fig.~\ref{fig:AdS4-sol}.
At each point of $\Sigma$ there is an $AdS_4$ and two 2-spheres, with independently varying radii.
The region $x\rightarrow\infty$ where the geometry becomes $AdS_5\times S^5$ is modeled in the Karch/Randall models in fig.~\ref{fig:KR-nongrav} by the $AdS_5$ region far away from the ETW brane.
The ETW brane itself can be understood as effective description for the remaining part of the 10d solution, i.e.\ the region around the 5-brane sources in fig.~\ref{fig:AdS4-sol}.
The intermediate holographic description, in which only the defect degrees of freedom are geometrized (description (b) above), corresponds to $AdS_4$ gravity in the region away from the $AdS_5\times S^5$ part coupled at the conformal boundary of $AdS_4$ to $\mathcal N=4$ SYM on a half space.
The 4d graviton has a mass, which, in the limit where the number of semi-infinite D3-branes is small, is set by the ratio of 4d and 3d central charges \cite{Bachas:2018zmb}.
We will modify these solutions by introducing black holes on the $AdS_4$ spaces, which leads to non-supersymmetric solutions of Type IIB that are asymptotic to the supersymmetric seed solutions and describe the dual QFTs at finite temperature.

The radiation region $R$ will be defined in the asymptotic $AdS_5\times S^5$ region at $x\rightarrow\infty$ in fig.~\ref{fig:AdS4-sol}, while the ``physical black hole'' corresponds to the region around the 5-brane sources.
The surfaces computing the entanglement entropy of the radiation region wrap both $S^2$'s and are anchored in the $AdS_5\times S^5$ region at a fixed value of the $AdS_4$ radial coordinate. 
For the non-gravitating baths we construct the HM surfaces explicitly at the time $t=0$ when their area is smallest.
The minimal surfaces can be described by specifying the $AdS_4$ radial coordinate $r$ as function of the coordinates on the Riemann surface $x$ and $y$.
The surfaces extend along the Riemann surface $\Sigma$, and either drop into the horizon in $AdS_4$ along a curve $x_h(y)$ (HM surfaces), or extend all the way to $x\rightarrow -\infty$, where they can close off smoothly before reaching the horizon in $AdS_4$ (island surfaces).

The extremality condition is a non-linear PDE on $\Sigma$. 
The boundary conditions will be derived from regularity of the induced metric on the minimal surface,
which will give a string theory justification for the use of Neumann boundary conditions at the ETW brane in the Karch/Randall models (other boundary conditions in 5d were discussed in \cite{Ghosh:2021axl}).
Solutions to the PDE are obtained numerically.
The class of $AdS_4\times S^2\times S^2\times \Sigma$ solutions is very broad, reflecting the breadth of brane configurations that can be realized with D3, D5 and NS5 branes.
We will choose representative solutions with $N_5$ D5-branes at $(x,y)=(0,0)$, $N_5$ NS5-branes at $(x,y)=(0,\frac{\pi}{2})$ and $2N_5K$ semi-infinite D3-branes. Studying more general solutions will be left for the future.

The 8d minimal surfaces can be visualized as 2d surfaces in the 3d space spanned by $\Sigma$ and the $AdS_4$ radial direction $r$,
with the horizon at some finite $r_h$. The conformal boundary of $AdS_4$ at $r\rightarrow\infty$ corresponds to the defect in fig.~\ref{fig:KR-nongrav}.
A sample of island and HM surfaces is shown in figs.~\ref{fig:islands}, \ref{fig:HM-surf}.
The island surfaces show distinct behavior near the 5-brane sources, which is discussed in sec.~\ref{sec:near-pole}.
The area differences between island surfaces and HM surfaces at $t=0$ are shown in fig.~\ref{fig:areadiff}. 
The results show that for radiation regions starting far in the bath (small $r$), the HM surface dominates at $t=0$.
The area of the HM surface grows in time and sets the initial growth of the entropy, but the entropy growth is bounded by the constant area of the island surface. 
This evades an information paradox and shows that the entropy follows a Page curve.

\medskip
\textbf{Critical angle:}
The analysis of  \cite{Geng:2020fxl} found a critical value for the tension/angle of the ETW brane ($\theta$ in fig.~\ref{fig:KR}), where the behavior of the island surfaces changes qualitatively.
The critical angle $\theta_c$ can be defined as follows:
At zero temperature, for an island surface anchored at a fixed point in the bath system, one can ask for the end point on the ETW brane as function of $\theta$. For $\theta>\theta_c$ this is a finite point.
As $\theta_c$ is approached, the end point on the ETW brane diverges towards the Poincar\'e horizon and below $\theta_c$ there are no more island minimal surfaces.

Remarkably, a similar phenomenon can be identified in 10d.
The angle $\theta$ in 5d is set by the tension of the ETW brane, which can be understood as a measure for the number of degrees of freedom represented by the ETW brane.
The relevant parameters in the 10d solutions considered here are the radius of the asymptotic $AdS_5\times S^5$ region, which is set by the number of semi-infinite D3-branes, and the number of D5 and NS5 branes on which the D3-branes terminate. 
The latter determines the 3d SCFT that $\mathcal N=4$ SYM is coupled to at the boundary of the half space.
One may expect that the brane angle in 5d captures the ratio of the number of D3-branes suspended between 5-branes and the number of semi-infinite D3-branes.
This is indeed the case: For island surfaces at zero temperature, with fixed anchor point in the $AdS_5\times S^5$ region, the end point at $x=-\infty$ is shown as function of $N_5/K$, which controls the ratio of suspended and semi-infinite D3-branes, in fig.~\ref{fig:crit-ang}.
The results indicate that there is a critical ratio at which the end point at $x=-\infty$ runs off towards the Poincar\'e horizon.
For black hole solutions with finite temperature this behavior is regulated (fig.~\ref{fig:crit-ang-T}), and island surfaces can be found  beyond the critical ratio.

\begin{figure}
	\begin{tikzpicture}
		\shade [ left color=blue! 0, right color=blue! 20] (-2.2,0)  rectangle (0,-2);
		\shade [ right color=blue! 0, left color=blue! 20] (0,0)  rectangle (2.2,-2);
		
		\draw[thick] (-2.2,0) -- (2.2,0);
		\draw[thick] (-2.2,-2) -- (2.2,-2);
		\draw[dashed] (-2.2,0) arc (160:200:85pt);
		\draw[dashed] (2.2,0) arc (20:-20:85pt);
		
		\node at (1.7,-0.6) {$\Sigma$};
		\node at (2.3,-2.3) {\small $x\rightarrow+\infty$};
		\node at (-2.3,-2.3) {\small $x\rightarrow-\infty$};
		
		\draw[thick] (0.2,-0.08) -- (0.2,0.08);
		\draw[thick] (-0.2,-0.08) -- (-0.2,0.08);
		\draw[thick] (0,-0.08) -- (0,0.08) node [anchor=south] {\small NS5};
		\draw[thick] (0.2,-2-0.08) -- (0.2,-2+0.08);
		\draw[thick] (-0.2,-2-0.08) -- (-0.2,-2+0.08);
		\draw[thick] (0,-1.92) -- (0,-2.08) node [anchor=north] {\small D5};
		
		\node at (-1,-1.75) {\small $y=0$};
		\node at (-1,-0.25) {\small $y=\frac{\pi}{2}$};
		
	\end{tikzpicture}
	\qquad\qquad
	\begin{tikzpicture}[y={(0cm,1cm)}, x={(0.707cm,0.707cm)}, z={(1cm,0cm)}, scale=1.1]
		\draw[white,fill=gray!100] (0,0,0.5) circle (1.5pt);
		\draw[white,fill=gray!100] (0,0,1.5) circle (2pt);
		\draw[white,fill=gray!100] (0,0,3.5) circle (1pt);
		
		\draw[thick] (0,-0.39,0) -- (0,1,0);
		\draw[thick] (0,-1,0) -- (0,-0.6,0);
		
		\draw[thick] (0,-0.41,1) -- (0,1,1);
		\draw[thick] (0,-1,1) -- (0,-0.61,1);
		
		\draw[thick] (0,-1,2) -- (0,1,2);
		
		\node at (0,0,2.5) {$\cdots$};
		
		\draw[thick] (0,-0.43,3) -- (0,1,3);
		\draw[thick] (0,-1,3) -- (0,-0.57,3);
		
		\foreach \i in {-0.075,-0.025,0.025,0.075}{ \draw (-1.1,\i,0.5) -- (0.65,\i,0.5);}
		\foreach \i in {-0.075,-0.025,0.025,0.075}{ \draw (0.76,\i,0.5) -- (1.1,\i,0.5);}
		
		\foreach \i in {-0.05,0,0.05}{ \draw (-1.1,\i,1.5) -- (0.65,\i,1.5);}
		\foreach \i in {-0.05,0,0.05}{ \draw (0.76,\i,1.5) -- (1.1,\i,1.5);}
		
		\foreach \i in {-0.025,0.025}{ \draw (-1.1,\i,3.5) -- (0.65,\i,3.5);}
		\foreach \i in {-0.025,0.025}{ \draw (0.76,\i,3.5) -- (1.1,\i,3.5);}
		
		\foreach \i in {-0.025,0,0.025}{ \draw (0,1.4*\i,0) -- (0,1.4*\i,1);}
		\foreach \i in {-0.05,-0.025,0,0.025,0.05}{ \draw (0,1.4*\i,1) -- (0,1.4*\i,2.05);}
		\foreach \i in {-0.045,-0.015,0.015,0.045}{ \draw (0,1.4*\i,2.95) -- (0,1.4*\i,3);}
		\foreach \i in {-0.015,0.015}{ \draw (0,1.4*\i,3) -- (0,1.4*\i,4);}
		
		\draw[thick] (0,-1,4) -- (0,1,4);
		
		\node at (-0.2,0,3.9) {\tiny D3};
		\node at (1.0,0.2,0.75) {\footnotesize D5};
		\node at (0,-1.25) {NS5};
	\end{tikzpicture}
	
	\caption{
		Left: $AdS_4\,{\times}\, S^2\,{\times}\, S^2\,{\times}\,\Sigma$ solutions for gravitating baths. The $AdS_5\times S^5$ region is closed off; the limits $x\rightarrow \pm\infty$ both lead to regular points in the internal space.
		This leaves the 3d conformal boundary of $AdS_4$, corresponding to the remaining point of the conformal boundary in fig.~\ref{fig:KR-grav}.
		Right: The associated brane configurations have no semi-infinite D3-branes, only D3-branes suspended between 5-branes.
		\label{fig:AdS4-sol-grav}}
\end{figure}
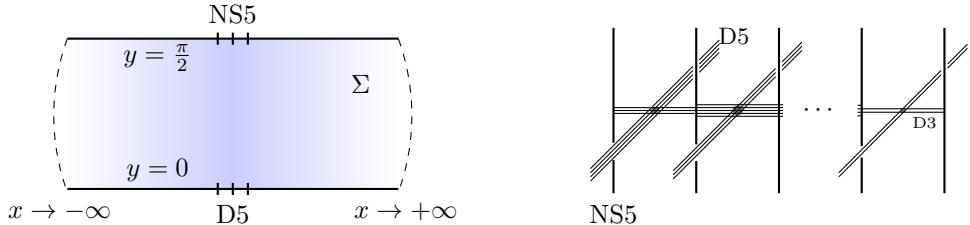

\medskip
\textbf{Gravitating baths:} 
For the description of a gravitating bath the asymptotic $AdS_5\times S^5$ region in fig.~\ref{fig:AdS4-sol} is closed off. 
This corresponds to removing the semi-infinite D3-branes from the brane setup, leaving only D3-branes suspended between D5 and NS5 branes (fig.~\ref{fig:AdS4-sol-grav}). 
This is captured in the 5d Karch/Randall models by the introduction of a second ETW brane. 
The 10d solutions are holographic duals for 3d $T_\rho^\sigma[SU(N)]$ SCFTs \cite{Assel:2011xz} and have massless 4d gravitons.
Closing off the $AdS_5\times S^5$ region removes the part in which the radiation region was defined,
and a minimal surface stretching from $x=-\infty$ to $x=+\infty$ now has to satisfy Neumann boundary conditions on both ends. This allows it to settle onto the black hole horizon and  leads to a constant entropy identical to the thermal entropy of the bath, 
 in line with the general arguments of \cite{Laddha:2020kvp,Raju:2020smc}.

One can instead consider minimal surfaces splitting the internal space, which are expected to compute non-geometric entanglement entropies (whose holographic interpretation was initiated in \cite{Mollabashi:2014qfa,Karch:2014pma}).
In the Karch/Randall models a ``left/right EE", represented by surfaces ending on the point where the two ETW branes meet in fig.~\ref{fig:KR-grav}, was found to exhibit Page curve behavior, and was interpreted as an internal entanglement entropy in \cite{Geng:2020fxl}.
The Type IIB solutions realize the dual of the defect as full 10d geometry, making them an ideal setting for studies of minimal surfaces separating degrees of freedom according to their representation in the internal space.

We consider surfaces wrapping the spatial part of $AdS_4$, both $S^2$'s, and a curve in $\Sigma$ which depends on the $AdS_4$ radial coordinate.
The surfaces are anchored at the conformal boundary of $AdS_4$ along a curve $x(y)$ in $\Sigma$ which separates the 5-brane sources and defines a split into black hole system and bath. 
Such surfaces may be expected to compute EEs associated with decompositions of the quiver diagram in the UV description of the dual 3d SCFT.
One again has to consider HM surfaces, extending through the horizon in $AdS_4$ into the thermofield double, and island surfaces which close off in one of the $x\rightarrow\pm\infty$ regions before reaching the horizon in $AdS_4$. These are 10d versions of the surfaces in fig.~\ref{fig:KR-grav}.
The class of $AdS_4\times S^2\times S^2\times \Sigma$ solutions that could be considered is again broad, and we focus on simple representatives.
We include two groups of D5-branes and two groups of NS5 branes, placed symmetrically at $x=\pm \delta$ on the boundary components of $\Sigma$. 
The separation of the 5-brane sources determines how the D3-branes in the associated brane configuration are suspended between the 5-branes.
Comparing to the Karch/Randall models in fig.~\ref{fig:KR-grav}, these particular 10d solutions correspond to two equal ETW brane angles.

Some 10d island surfaces are shown in fig.~\ref{fig:LRcrit2}. 
The corresponding HM surface is described by $x=0$ and a time-dependent embedding in the $AdS_4$ part of the geometry.
The difference in areas between island and HM surfaces at $t=0$ is shown in fig.~\ref{fig:LRcrit1b}.
We find that for $\delta$ above a ``Page value" $\delta_P$ the HM surface initially dominates at $t=0$.
The entropy growth indicated by the HM surfaces is bounded by the constant area of the island surfaces, leading again to Page curves, shown in fig.~\ref{fig:page}.
A second distinguished value for $\delta$ can be seen in fig.~\ref{fig:LRcrit1a}:
at a critical value $\delta_c$ the cap-off point of the island surface at $x=-\infty$ diverges towards the conformal boundary of $AdS_4$,
and no island minimal surfaces are found for $\delta<\delta_c$.
The numerical results suggest that $\delta_c$ is slightly smaller than $\delta_P$, though we leave the possibility that the difference could be a numerical artifact.
In the small (and possibly empty) range $\delta_c<\delta<\delta_P$ the island surfaces are found to dominate already at $t=0$, leading to a flat entropy curve.
These results bear striking resemblance with critical and Page angles found in the Karch/Randall models in \cite{Geng:2020fxl}, suggesting that the ETW brane angles capture aspects of how the 5-brane sources are distributed on $\Sigma$ in~10d.

In the regime where no island minimal surfaces were found in the 5d Karch/Randall models in \cite{Geng:2020fxl}, ``tiny island"
limiting surfaces, which degenerate to an infinitesimal segment at the defect in fig.~\ref{fig:KR-grav}, were found to dominate and limit the entropy growth indicated by the HM surface.
In 10d we find that similar tiny island surfaces connecting the $x=0$ locus to $x=\pm\infty$ arise for $\delta<\delta_c$.

\medskip 
\textbf{Outline:} 
The main part is organized as follows.
The 10d supergravity solutions are introduced in sec.~\ref{sec:IIBsol}.
In sec.~\ref{sec:surfaces} the ansatz for extremal surfaces is discussed along with the extremality and boundary conditions and the behavior near the 5-branes.
The method for constructing minimal surfaces is summarized in sec.~\ref{sec:numerics}.
Island surfaces and Page curves are discussed for non-gravitating baths  in sec.~\ref{sec:islands} and for gravitating baths in sec.~\ref{sec:grav-bath}. We close with a brief outlook in sec.~\ref{sec:outlook}.

\section{Type IIB supergravity solutions}\label{sec:IIBsol}

The general local form of the $AdS_4\times S^2\times S^2\times \Sigma$ solutions that will be used here was constructed in \cite{DHoker:2007zhm,DHoker:2007hhe}. 
For the study of minimal surfaces we will only need the geometry, which is a warped product of $AdS_4$ and two 2-spheres, $S_1^2$ and $S_2^2$, over a Riemann surface $\Sigma$.
For the solutions of interest here $\Sigma$ can be taken as a strip, 
\begin{align}
	\Sigma&=\lbrace z\in\mathds{C}\,\vert\, 0\leq \Im(z)\leq \pi/2\rbrace~.
\end{align}
On each of the boundary components of the strip one of the $S^2$'s closes off smoothly, so that the 10d geometry has no boundary.
Depending on the nature of the points at infinity, solutions for different types of field theories can be constructed: 
Janus solutions, dual to interface CFTs, can be realized if the points $\Re(z)\rightarrow \pm\infty$ both correspond to asymptotic $AdS_5\times S^5$ regions.
Solutions with one asymptotic region closed off were constructed in \cite{Aharony:2011yc} and are dual to BCFTs.
Duals for 3d SCFTs were constructed in \cite{Assel:2011xz} by closing both asymptotic $AdS_5\times S^5$ regions.

The solutions are generally parametrized by two harmonic functions $h_1$, $h_2$ on $\Sigma$.
The Einstein-frame metric takes the form
\begin{align}\label{eq:10d-metric}
	ds^2&=f_4^2 ds^2_{4}+f_1^2 ds^2_{S_1^2}+f_2^2 ds^2_{S_2^2}+4\rho^2 |dz|^2~,
\end{align}
where $ds^2_{4}$ and $ds^2_{S_i^2}$ are line elements of unit-radius $AdS_4$ and $S^2$, respectively.
The coefficient functions are given by
\begin{align}
	f_4^8&=16\frac{N_1N_2}{W^2}~, & f_1^8&=16h_1^8\frac{N_2 W^2}{N_1^3}~, & f_2^8&=16 h_2^8 \frac{N_1 W^2}{N_2^3}~,
	&
	\rho^8&=\frac{N_1N_2W^2}{h_1^4h_2^4}~,
\end{align}
where
\begin{align}
	W&=\partial\bar\partial (h_1 h_2)~, & N_i &=2h_1 h_2 |\partial h_i|^2 -h_i^2 W~.
\end{align}
The expressions for the fluxes and dilaton will not be needed here; they can be found in \cite{DHoker:2007zhm,DHoker:2007hhe,Aharony:2011yc,Assel:2011xz}.

Based on this local form broad classes of supergravity solutions can be constructed which describe D3-branes intersecting, ending on, or suspended between D5 and NS5 branes.
For the realization of Karch/Randall models with gravitating and non-gravitating baths we will employ representative solutions dual to BCFTs and 3d SCFTs, noting that more general solutions could be considered.
The form of the harmonic functions $h_1$, $h_2$ is
\begin{align}
	h_1&=\frac{\pi \alpha^\prime}{4} K e^z-\frac{\alpha^\prime}{4} \sum_{a}N_{\rm D5}^{(a)}\ln\tanh\left(\frac{z-\delta_a}{2}\right)+\rm{c.c.}
	\nonumber\\
	h_2&=-\frac{i \pi \alpha^\prime}{4} K e^z-\frac{\alpha^\prime}{4}\sum_b N_{\rm NS5}^{(b)}\ln\tanh\left(\frac{i\pi}{4}-\frac{z-\delta_b}{2}\right)+\rm{c.c.}
\end{align}
The solutions describe semi-infinite D3-branes ending on D5-branes and NS5-branes which have additional D3-branes suspended between them.
The number of semi-infinite D3-branes is controlled by $K$; for $K=0$ the solutions describe D3-branes suspended between D5 and NS5 branes.
Groups of D5/NS5 branes are represented by the poles of $\partial h_1$/$\partial h_2$ on the boundary of $\Sigma$.
The specific brane configuration can be characterized in terms of linking numbers, which are encoded in the distribution of the 5-brane sources on $\Sigma$ \cite{Aharony:2011yc,Assel:2011xz}.
For $K\neq 0$ an $AdS_5\times S^5$ region emerges at $\Re(z)\rightarrow + \infty$, with $\Re(z)$ becoming the radial coordinate of $AdS_5$ in $AdS_4$ slicing and $\Im(z)$ becoming an angular coordinate on $S^5$.
For $K=0$ the limit $\Re(z)\rightarrow\infty$ leads to a regular point in the internal space.
The limit $\Re(z)\rightarrow -\infty$ leads to a regular point in both cases.

\begin{figure}
	\subfigure[][]{\label{fig:brane-non-grav}
	\begin{tikzpicture}[y={(0cm,1cm)}, x={(0.707cm,0.707cm)}, z={(1cm,0cm)}, scale=1.1]
	\draw[gray,fill=gray!100,rotate around={-45:(0,0,2)}] (0,0,2) ellipse (1.8pt and 3.5pt);
	\draw[gray,fill=gray!100] (0,0,0) circle (1.5pt);
	
	\foreach \i in {-0.05,0,0.05}{ \draw[thick] (0,-1,\i) -- (0,1,\i);}

	\foreach \i in {-0.075,-0.025,0.025,0.075}{ \draw (-1.1,\i,2) -- (1.1,\i,2);}
		
	\foreach \i in {-0.045,-0.015,0.015,0.045}{ \draw (0,1.4*\i,0) -- (0,1.4*\i,2+\i);}
	\foreach \i in  {-0.075,-0.045,-0.015,0.015,0.045,0.075}{ \draw (0,1.4*\i,2+\i) -- (0,1.4*\i,4);}
	
	\node at (-0.18,-0.18,3.4) {\scriptsize $2N_5 K$};
	\node at (1.0,0.3,2) {\scriptsize $N_5$ D5};
	\node at (0,-1.25) {\footnotesize $N_5$ NS5};
	\node at (0.18,0.18,0.9) {{\scriptsize $N_5 K+\tfrac{N_5^2}{2}$}};
\end{tikzpicture}
}
\hskip 20mm
\subfigure[][]{\label{fig:brane-grav}
		\begin{tikzpicture}[y={(0cm,1cm)}, x={(0.707cm,0.707cm)}, z={(1cm,0cm)}, scale=1.1]
		\draw[gray,fill=gray!100] (0,0,-0.5) circle (1.8pt);
		\draw[gray,fill=gray!100] (0,0,1) ellipse (1.8pt and 3pt);
		\draw[gray,fill=gray!100,rotate around={-45:(0,0,2.5)}] (0,0,2.5) ellipse (1.8pt and 3.5pt);
		\draw[gray,fill=gray!100] (0,0,4) circle (1.8pt);				
		
		\foreach \i in {-0.05,0,0.05}{ \draw[thick] (0,-1,-0.5+\i) -- (0,1,-0.5+\i);}
		\foreach \i in {-0.05,0,0.05}{ \draw[thick] (0,-1,1+\i) -- (0,1,1+\i);}

		\foreach \i in {-0.075,-0.025,0.025,0.075}{ \draw (-1.1,\i,2.5) -- (1.1,\i,2.5);}
		\foreach \i in {-0.075,-0.025,0.025,0.075}{ \draw (-1.1,\i,4) -- (1.1,\i,4);}
		
		\foreach \i in {-0.03,0,0.03}{ \draw (0,1.4*\i,-0.5) -- (0,1.4*\i,1);}
		\foreach \i in {-0.075,-0.045,-0.015,0.015,0.045,0.075}{ \draw (0,1.4*\i,1) -- (0,1.4*\i,2.5+\i);}
		\foreach \i in {-0.03,0,0.03}{ \draw (0,1.4*\i,2.5) -- (0,1.4*\i,4);}

		\node at (0,-1.25,-0.5) {\footnotesize $\tfrac{N_5}{2}$ NS5};
		\node at (0,-1.25,1) {\footnotesize $\tfrac{N_5}{2}$ NS5};
		\node at (1.0,0.35,2.5) {\scriptsize  $\tfrac{N_5}{2}$ D5};
		\node at (1.0,0.35,4) {\scriptsize  $\tfrac{N_5}{2}$ D5};
		\node at (0.22,0.22,1.75) {{\scriptsize $\tfrac{N_5^2}{2}$}};
		\node at (0,0.3,0.25) {{\scriptsize $\tfrac{N_5^2}{4}\Delta$}};
		\node at (0,0.3,3.5) {{\scriptsize $\tfrac{N_5^2}{4}\Delta$}};
	\end{tikzpicture}
}
\caption{Brane configurations for representative non-gravitating bath solutions (left) and gravitating bath solutions (right). 
	Hanany-Witten transitions can be used to make the 3d quiver gauge theories more apparent, as in figs.~\ref{fig:AdS4-sol}, \ref{fig:AdS4-sol-grav}. The numbers of D3-branes  on the right are controlled by $\delta$ through $\Delta=\frac{1}{2}+\frac{2}{\pi}\arctan e^{2\delta}$.
}
\end{figure}
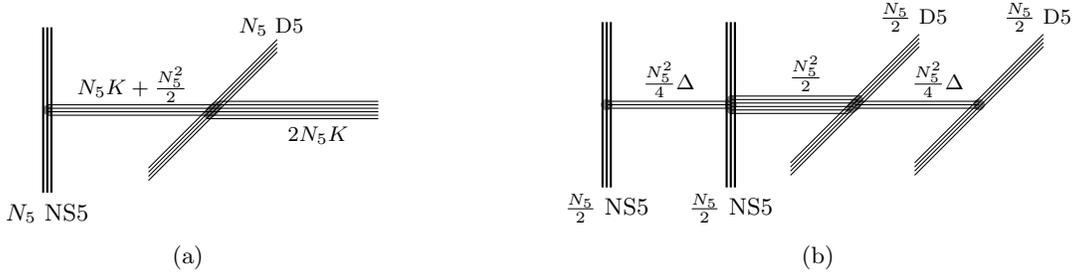

We discuss the concrete solutions that will be used below first and briefly comment on the more general picture and dual field theories afterwards.
The solutions we will study for non-gravitating baths are dual to $\mathcal N=4$ SYM on a half space coupled to 3d $T_\rho^\sigma[SU(N)]$ theories on the boundary.
They are given by $h_{1/2}$ of the form
\begin{align}\label{eq:h1h2-BCFT}
	h_1&=\frac{\pi \alpha^\prime}{4} K e^z-\frac{\alpha^\prime}{4}N_5\ln\tanh\left(\frac{z}{2}\right)+\rm{c.c.}
	\nonumber\\
	h_2&=-\frac{i\pi\alpha^\prime}{4}K e^z-\frac{\alpha^\prime}{4}N_5\ln\tanh\left(\frac{i\pi}{4}-\frac{z}{2}\right)+\rm{c.c.}
\end{align}
The radii of $AdS_5$ and $S^5$ in the $AdS_5\times S^5$ region at $\Re(z)\rightarrow\infty$ are set by $L^4=8\pi{\alpha^\prime}^2N_5K$.
The asymptotic string coupling is $\lim_{x\rightarrow\infty}e^\phi=1$.
These solutions are string theory realizations of the Karch/Randall models with one ETW brane (fig.~\ref{fig:KR-nongrav}): the asymptotic region at $\Re(z)\rightarrow\infty$ corresponds to the $AdS_5$ part, while the region with the NS5/D5 sources is the string theory version of the ETW brane itself.
The brane configuration involves $2N_5K$ semi-infinite D3-branes ending on a combination of $N_5$ D5-branes and $N_5$ NS5-branes (fig.~\ref{fig:brane-non-grav}).
$N_5K$ D3-branes end on the D5 branes and $N_5K$ D3-branes end on the NS5-branes, and there are in addition $N_5^2/2$ D3-branes suspended between the D5 and NS5 branes.

The solutions for gravitating baths that will be considered below are holographic duals for 3d $T_\rho^\sigma[SU(N)]$ SCFTs.
The functions $h_1$ and $h_2$ are given by
\begin{align}\label{eq:h1h2-3d-grav}
	h_1&=-\frac{\alpha^\prime}{4}\frac{N_5}{2}\left[\ln\tanh\left(\frac{z-\delta}{2}\right)+
\ln	\tanh\left(\frac{z+\delta}{2}\right)\right]+\rm{c.c.}
	\nonumber\\
	h_2&=-\frac{\alpha^\prime}{4}\frac{N_5}{2}\left[\ln\tanh\left(\frac{i\pi}{4}-\frac{z-\delta}{2}\right)
	+\ln\tanh\left(\frac{i\pi}{4}-\frac{z+\delta}{2}\right)\right]+\rm{c.c.}
\end{align}
These solutions describe $N_5^2/2$ D3-branes suspended between two groups of D5-branes and two groups of NS5-branes, with $N_5/2$ 5-branes in each group.
There are no semi-infinite D3-branes and the asymptotic $AdS_5\times S^5$ region at $\Re(z)\rightarrow \infty$ is closed off.
The limits $\Re(z)\rightarrow \pm\infty$ both correspond to regular points in the internal space.
The 5-brane groups are represented in the supergravity solutions by sources with $N_5/2$ D5 and $N_5/2$ NS5-branes, respectively, at $z=\pm\delta$ and $z=\pm\delta+i\pi/2$. The parameter $ \delta$ determines how the D3-branes terminate on the D5 and NS5 branes (fig.~\ref{fig:brane-grav}); for $\delta=0$ the numbers of D3-branes terminating on each group of 5-branes are equal.
The dual 3d SCFTs are special cases of the theories discussed in sec.~5.3 of \cite{Coccia:2020wtk}.
Comparing to the 5d Karch/Randall models, the closing off of the asymptotic $AdS_5\times S^5$ region corresponds to the introduction of the second ETW brane in fig.~\ref{fig:KR-grav}. The entire 10d solution corresponds to the remaining wedge of $AdS_5$ in fig.~\ref{fig:KR-grav}.

The solutions (\ref{eq:h1h2-BCFT}) and (\ref{eq:h1h2-3d-grav}) are invariant under S-duality (exchange of $h_1$ and $h_2$ combined with $z\rightarrow \frac{i\pi}{2}-z$),
reflecting that the associated brane configurations are invariant under S-duality (in fig.~\ref{fig:brane-non-grav} up to Hanany-Witten transitions).
This will be useful below. 
From now on we set $\alpha^\prime=1$.

Solutions with more general arrangements of 5-brane sources (poles in $\partial h_{1/2}$) and no asymptotic $AdS_5\times S^5$ region describe configurations of D3-branes suspended between D5 and NS5 branes that can be characterized by two Young tableaux $\rho$ and $\sigma$, which determine how precisely the D3-branes terminate on the 5-branes.
The general relation between the distribution of the 5-brane sources on the boundary of $\Sigma$ and the Young tableaux $\rho$ and $\sigma$ can be found in  \cite{Assel:2011xz}.
The brane configurations engineer 3d $\mathcal N=4$ quiver gauge theories, and the supergravity solutions are dual to their IR fixed points.
For solutions with $AdS_5\times S^5$ region and semi-infinite D3-branes the dual field theory is $\mathcal N=4$ SYM on a half space coupled to a 3d $T_\rho^\sigma[SU(N)]$ SCFT on the boundary \cite{Aharony:2011yc}. 
The free energies obtained holographically were matched to field theory computations using supersymmetric localization for the former in \cite{Assel:2012cp}, \cite{Coccia:2020wtk} and for the latter in \cite{,VanRaamsdonk:2020djx}.

\subsection{Finite temperature}
For each  $AdS_4\times S^2\times S^2\times\Sigma$ solution one may replace $AdS_4$ by a finite temperature black hole and still obtain a solution to the Type IIB supergravity field equations:
To verify the field equations one only needs that the 4d space is Einstein with negative curvature.
This is true for the $AdS_4$ black hole metrics we will use, so that replacing $AdS_4$ by a black hole yields non-supersymmetric solutions which  asymptotically approach the supersymmetric seed solution.
From a more general perspective, the $AdS_4\times S^2\times S^2\times\Sigma$ solutions are in the class for which \cite{Gauntlett:2007ma} conjecture that a  consistent truncation exists. 
Having a consistent truncation to 4d gauged supergravity would allow to uplift more general 4d solutions to 10d, but this is not needed for our purposes here.

To introduce finite temperature, we replace the $AdS_4$ metric in (\ref{eq:10d-metric}) by 
the $AdS_4$ black hole metric
\begin{align}\label{eq:ds2-AdS4-T}
	ds_4^2&=\frac{dr^2}{b(r)}+e^{2r}\left(-b(r)dt^2+ds^2_{\RR^2}\right)~,
	&
	b(r)&=1-e^{3(r_h-r)}~.
\end{align}
The horizon is at $r=r_h$, the conformal boundary at $r\rightarrow\infty$.
It will be convenient to also introduce the tortoise coordinate $u$ by
\begin{align}\label{eq:tortoise}
	du&=\frac{dr}{\sqrt{b(r)}}~, & u&=\frac{2}{3}\cosh^{-1}\left(e^{\frac{3}{2}(r-r_h)}\right)~.
\end{align}
The range $u\in\RR^+$ corresponds to the exterior region covered by the original coordinate $r$, with the horizon at $u=0$. The metric becomes
\begin{align}
	ds^2_4&=du^2+e^{2r_h}\cosh^{4/3}\left(\frac{3u}{2}\right)\left[-\tanh^2\left(\frac{3u}{2}\right)dt^2+ds^2_{\RR^2}\right]~.
\end{align}
From the CFT perspective replacing $AdS_4$ by a planar black hole corresponds to adding a finite temperature for $\mathcal N=4$ SYM on $AdS_4$ for solutions with an $AdS_5\times S^5$ region.
The black hole solutions without $AdS_5\times S^5$ region are dual to 3d $T_\rho^\sigma[SU(N)]$ SCFTs at finite temperature.

\section{Extremal surfaces}\label{sec:surfaces}

In this section we discuss the embedding ansatz for the surfaces that will be used for the entanglement entropy computations,
the extremality and boundary conditions, and the behavior near the 5-brane sources.

\subsection{Island surfaces}

The surfaces of interest are 8d minimal surfaces in the 10d geometry (\ref{eq:10d-metric}) that wrap both $S^2$'s, (part of) the Riemann surface $\Sigma$, and a part of the $AdS_4$ black hole geometry.
For the $AdS_4$ black hole we choose coordinates (\ref{eq:ds2-AdS4-T}), such that the 10d metric is given by (\ref{eq:10d-metric}) with
\begin{align}
		ds_4^2&=\frac{dr^2}{b(r)}+e^{2r}\left(-b(r)dt^2+ds^2_{\RR^2}\right)~.
\end{align}
The surfaces can be described by specifying the $AdS_4$ radial coordinate $r$ for any given point of $\Sigma$.
On $\Sigma$ we introduce real coordinates
\begin{align}
	z=x+iy~,
\end{align}
with $x\in\mathds{R}$ and $0\leq y\leq\frac{\pi}{2}$.
The embeddings are thus described by a single embedding function
\begin{align}
	r=r(x,y)~.
\end{align}
The induced metric on the surface reads
\begin{align}\label{eq:ind-met}
	ds^2_\gamma&=e^{2r}f_4^2ds^2_{\RR^2}+f_1^2ds^2_{S_1^2}+f_2^2ds^2_{S^2_2}+4\rho^2 (dx^2+dy^2)
	+\frac{f_4^2}{b(r)}\left(dx\, \partial_x r +dy \partial_y r \right)^2~.
\end{align}
The area of a general surface of this form is given by $A=V_{\RR^2}V_{S_1^2\times S_2^2}S_\gamma$, with
\begin{align}
	S_\gamma&=4\int dx dy \,e^{2r}f_4^2f_1^2f_2^2\rho^2\sqrt{1+\frac{f_4^2}{4b(r)\rho^2}\left((\partial_x r)^2+(\partial_y r)^2\right)}~.
\end{align}
The combinations of metric functions appearing in this expression are given by
\begin{align}\label{eq:fg-eval}
	f_4^2 f_1^2f_2^2\rho^2&=8\left|h_1 h_2 W\right|~,
	&
	\frac{f_4^2}{\rho^2}&=2\left|\frac{h_1h_2}{W}\right|~.
\end{align}
With these expressions the area simplifies to
\begin{align}\label{eq:S}
	S_\gamma&=32\int dx dy \,e^{2r}\left|h_1 h_2 W\right|\sqrt{1+\frac{1}{2b(r)}\left|\frac{h_1 h_2}{W}\right| (\nabla r)^2}~.
\end{align}
Since $4W=\Delta(h_1h_2)$, the area depends on $h_1$ and $h_2$ only through the combination $h_1h_2$.

The extremality condition resulting from variation of $S_\gamma$ (with $S_\gamma=\int L_\gamma$) can be written as
\begin{align}\label{eq:eom-fg}
	0\stackrel{!}{=}	\frac{\delta L_\gamma}{L_\gamma}&=
	\frac{1}{1+g(\nabla r)^2}\left[2-\nabla(g\nabla r)+\frac{1}{2}g\nabla r\cdot \nabla\ln\left(\frac{1+g(\nabla r)^2)}{b(r) f^2}\right)\right]~,
\end{align}
where $\nabla$ is the covariant derivative with respect to the metric on $\Sigma$ and
\begin{align}
	f&=|h_1 h_2 W|~, & g&=\frac{1}{2b(r)}\left|\frac{h_1 h_2}{W}\right|~.
\end{align}
The dependence on $r$ itself drops out for zero temperature, i.e.\ when $b(r)=1$.
If $r(x,y)$ is a solution to the extremality condition at zero temperature then so is $r(x,y)+c$ with a constant $c$, with different asymptotic values at $x\rightarrow \pm\infty$; this reflects the defect conformal symmetry.

\subsection{Boundary conditions}\label{sec:bc}

We now discuss the boundary conditions for surfaces extending along $\Sigma$, starting with the two boundary components of the strip at $y=0$ and $y=\frac{\pi}{2}$.
Near $y=0$ the sphere $S_1^2$, collapses, with $f_1^2\sim 4y^2 \rho^2$ so that the background has no conical singularity in the space parametrized by $y$ and $S_1^2$.
The induced metric (\ref{eq:ind-met}) near $y=0$ consequently takes the form
\begin{align}
	ds^2_\gamma&\approx e^{2r}f_4^2ds^2_{\RR^2}+f_2^2ds^2_{S^2_2}+4\rho^2 \left( dx^2+dy^2+y^2ds^2_{S_1^2}\right)+\frac{f_4^2}{b(r)}\left(dx\, \partial_x r +dy \partial_y r \right)^2~.
\end{align}
The contribution proportional to $(\partial_y r)^2 dy^2$ threatens to introduce a conical singularity in the $(y,S_1^2)$ part of the induced metric on the surface. A smooth metric is obtained with the Neumann boundary condition $\partial_y r\vert_{y=0}=0$.
The reasoning for the second boundary component, where $S_2^2$ collapses, is analogous. We conclude
\begin{align}\label{eq:Neumann-bc-y}
	\partial_y r(x,y)\big\vert_{y=0}&=0~, & \partial_y r(x,y)\big\vert_{y=\frac{\pi}{2}}&=0~.
\end{align}

For $x\rightarrow -\infty$ the space closes off smoothly; the limit corresponds to a single regular point on the boundary of $\Sigma$. For the surface to be smooth,  $\lim_{x\rightarrow -\infty}r(x,y)$ should be independent of $y$.
The asymptotic behavior of the metric functions, with coordinate $v=2e^{x}$ and $v\rightarrow 0$, is given by (see (3.15) of \cite{Assel:2011xz})
\begin{align}
	f_4^2&\approx L^2~, & f_1^2&\approx 4\sin^2\!y\,\rho^2~, & f_2^2&\approx 4\cos^2\!y\,\rho^2~, & 4\rho^2&\approx L^2v^2~.
\end{align}
The induced metric on the minimal surface becomes (noting that $\partial_y r\rightarrow 0$)
\begin{align}
	ds^2_\gamma&\approx L^2\left[ e^{2r}ds^2_{\RR^2}+dv^2+v^2\left(dy^2+\sin^2\!y\, ds^2_{S_1^2}+\cos^2\!y\,ds^2_{S_2^2}\right)+(\partial_x r)^2\frac{dv^2}{v^2}\right]~.
\end{align}
The part in the round bracket is the line element for $S^5$, and a smooth $\RR^8$ with no conical singularity is obtained if
\begin{align}\label{eq:bc-x-minus}
	\lim_{x\rightarrow-\infty}e^{-x}\partial_x r(x,y)=0~.
\end{align}
The conditions (\ref{eq:Neumann-bc-y}) and (\ref{eq:bc-x-minus}) are the 10d analog of the Neumann boundary conditions imposed at the ETW brane in the 5d Karch/Randall models.

The nature of the limit $x\rightarrow +\infty$ is different for the solutions in (\ref{eq:h1h2-BCFT}) for a non-gravitating bath, where an $AdS_5\times S^5$ region emerges in this limit, compared to the solution (\ref{eq:h1h2-3d-grav}) for a gravitating bath.
For the latter the limits $x\rightarrow \pm \infty$ both lead to regular boundary points, and the boundary condition at $x\rightarrow+\infty$ is given by (\ref{eq:bc-x-minus}) with $x\rightarrow -x$.
For the former, with the emerging $AdS_5\times S^5$ region, a Dirichlet condition anchoring the surface is imposed instead. 
The general form is 
\begin{align}
	\lim_{x\rightarrow+\infty} r(x,y)&=r_0(y)~.
\end{align}
The form of $r_0(y)$ can be determined by considering global $AdS_5\times S^5$, corresponding to $h_1=\cosh z+\rm{c.c.}$ and $h_2=-i \sinh z+\rm{c.c.}$ 
In that case $|h_1h_2/W|=2\cosh^2(x)$, which is independent of $y$.
As a result one can find extremal surfaces with no dependence on $y$, which is an angular coordinate on $S^5$.
For more general solutions the  boundary condition in the asymptotic $AdS_5\times S^5$ region at $x\rightarrow\infty$ therefore is that $r(x,y)$ should become independent of $y$ and satisfy a Dirichlet condition with $r_0(y)=r_R$. In summary,
\begin{align}\label{eq:Dirichlet}
	\lim_{x\rightarrow+\infty}r(x,y)&=r^{}_R\qquad \text{for (\ref{eq:h1h2-BCFT}),}
	&
	\lim_{x\rightarrow+\infty}e^{+x}\partial_x r(x,y)&=0 \qquad \text{for (\ref{eq:h1h2-3d-grav}).}
\end{align}

\subsection{Near-pole behavior}\label{sec:near-pole}

At zero temperature the minimal surfaces will show distinct behavior near the 5-brane sources, and cap off there.\footnote{This differs from the behavior of the spherical entangling surface centered on the defect studied in \cite{VanRaamsdonk:2020djx}, which has a simple universal embedding which is insensitive to the 5-brane sources.}
In this section we will discuss this behavior analytically, using the form of the supergravity solutions near the 5-brane sources.
At finite temperature the behavior near the 5-brane sources will be regulated by the horizon.

To discuss the behavior near a pole at $z=z_0$ it is convenient to introduce coordinates centered on the pole, 
$z=z_0+R e^{i\varphi}$ for $z_0$ on the real line and $z=z_0-Re^{i\varphi}$ for $\Im(z_0)=\pi/2$.
The combinations that appear in the area functional (\ref{eq:S}) behave at zero temperature as follows,
\begin{align}
	f=|h_1h_2 W|&\approx f_0\sin^2(\varphi)(-\ln R)~,
	&
	g=\frac{1}{2}\left|\frac{h_1h_2}{W}\right|&\approx -R^2\ln R~,
\end{align}
where $f_0$ is a constant which depends on the solution under consideration. 
The value of $f_0$ will not be relevant, since the extremality condition (\ref{eq:eom-fg}) is invariant under constant rescalings of $f$.

To discuss the near-pole behavior it is convenient to drop the overall factor in the extremality condition (\ref{eq:eom-fg}) and use the condition in the form
\begin{align}\label{eq:eom-nb}
	0&=2-\nabla\left(g\nabla r\right)+\frac{1}{2}g\nabla r\cdot \nabla\ln\left(\frac{1+g(\nabla r)^2}{f^2}\right)~.
\end{align}
The two non-trivial terms on the right hand side are generically of the same order, noting that $\nabla \ln(\ldots)=\mathcal O(1/R)$.
A scaling analysis suggests to take $\nabla r=\mathcal O(1/(R\ln R))$ and make an ansatz
\begin{align}
	r(R,\varphi)&= r_0\ln(-\ln R)+\frac{r_1(\varphi)}{\ln R}+\ldots
\end{align}
where the ellipsis denotes regular and subleading terms.
The leading non-trivial order in the extremality condition (\ref{eq:eom-nb}) then is its finite part. 
The near-pole solution without divergences in $\varphi$ is given by $r_0=-1$ and $r_1$ constant.
In summary, the behavior of the embedding near a 5-brane source at $z=z_0$ is given by
\begin{align}\label{eq:r-near-pole}
	r(z,\bar z)&= -\ln(-\ln |z-z_0|)+\ldots~.
\end{align}
Since $\lim_{z\rightarrow z_0}r(z,\bar z)=-\infty$, the minimal surface drops into the Poincar\'e horizon at the source.
At the point $z_0$ the background geometry is singular, as appropriate for a solution near a 5-brane source, and we do not impose additional regularity conditions for the minimal surface.

\subsection{HM surfaces}

We will focus on the non-gravitating bath solutions (\ref{eq:h1h2-BCFT}) for the discussion of HM surfaces; those for the gravitating bath solutions will be discussed in sec.~\ref{sec:grav-bath}.
We use the tortoise coordinate $u$ defined in (\ref{eq:tortoise}) and parametrize the embedding at $t=0$ in terms of $x(u,y)$ instead of $r(x,y)$.
The minimal surfaces range in $u$ from the value enforced by the Dirichlet boundary condition (\ref{eq:Dirichlet}) through the horizon into the thermofield double.
We focus on surfaces anchored at the same point $r_R$ in the thermofield double, which are symmetric with respect to reflection across $u=0$ at $t=0$.
So we can restrict to $u\geq 0$ to find the embeddings.
From that perspective the HM surfaces end on the horizon at $u=0$ along a curve $x_h(y)$ which is determined by the extremality condition.

The induced metric with the tortoise coordinate $u$ and the parametrization $x(u,y)$ becomes 
\begin{align}
	ds^2=\,&e^{2r_h}\cosh^{4/3}\left(\frac{3u}{2}\right)f_4^2ds^2_{\mathds{R}^2}+f_1^2 ds^2_{S_1^2}+f_2^2ds^2_{S_2^2}+
	\left[f_4^2+4\rho^2(\partial_u x)^2\right]du^2
	\nonumber\\ &
	+4\rho^2 \left[dy^2\left(1+(\partial_y x)^2\right)+(\partial_u x)(\partial_y x)(du\, dy+dy\,du)\right]~.
\end{align}
The area evaluated using (\ref{eq:fg-eval}) becomes
\begin{align}\label{eq:HM-area}
	S&=32\int du dy\, e^{2r_h}\cosh^{4/3}\left(\frac{3u}{2}\right)|h_1 h_2W|\sqrt{\frac{1}{2}\left|\frac{h_1 h_2}{W}\right|\left(1+(\partial_y x)^2\right)+(\partial_u x)^2}~.
\end{align}

For the boundary conditions we start with the boundaries of $\Sigma$ at $y=0$ and $y=\frac{\pi}{2}$.
Having no conical singularities at $y=0,\frac{\pi}{2}$ leads to the Neumann boundary conditions
\begin{align}\label{eq:Neumann-bc-y-HM}
	\partial_y x(u,y)\big\vert_{y=0}&=0~, &\partial_y x(u,y)\big\vert_{y=\frac{\pi}{2}}&=0~,
\end{align}
analogously to the arguments for  (\ref{eq:Neumann-bc-y}) before.
The Dirichlet condition which anchors the surface, given in (\ref{eq:Dirichlet}) for the parametrization $r(x,y)$, here becomes
\begin{align}
	\lim_{u\rightarrow u(r_R)} x(u,y)=\infty~.
\end{align}
On the other end the surface should intersect the horizon and end from the one-sided perspective at $u=0$.
The symmetry under reflection across $u=0$ leads to the Neumann condition
\begin{align}
	\partial_u x(u,y)\vert_{u=0}&=0~.
\end{align}
This condition also ensures that boundary terms in the variation of the area  at $u=0$ vanish. 

\section{Solving for minimal surfaces}\label{sec:numerics}

To summarize, the extremality conditions are non-linear second-order PDEs on the strip $\Sigma=\lbrace x+iy\vert x\in\mathds{R}, 0\leq y\leq\frac{\pi}{2}\rbrace$, with Neumann boundary conditions at $y=0$ and $y=\frac{\pi}{2}$.
The domain and boundary conditions in the $x$ direction depend on the background solution and type of surface under consideration.
The solutions are expected to be smooth, except for the locations on the two boundary components at $y\in\lbrace 0,\frac{\pi}{2}\rbrace$ where the D5/NS5 sources are in (\ref{eq:h1h2-BCFT}) and (\ref{eq:h1h2-3d-grav}).

To solve these PDEs numerically we start with a trial surface satisfying the boundary conditions and let it dynamically settle on a minimal area configuration.
To this end an auxiliary external time parameter $\tau$ is introduced, and the embedding, say for island surfaces, is described by a $\tau$-dependent function $r(x,y,\tau)$.
The $\tau$-evolution for $r(x,y,\tau)$ is chosen as
\begin{align}\label{eq:r-tau}
	\partial_\tau r(x,y,\tau)&=-L_\gamma^{-1}\frac{\delta L_\gamma}{\delta r(x,y,\tau)}~,
\end{align}
where $L_\gamma$ is the volume element of the surface in (\ref{eq:S}).
This exerts a force on the embedding in the direction in which the area decreases.
The right hand side is given by (\ref{eq:eom-fg}) with $r(x,y)$ replaced by $r(x,y,\tau)$.

To numerically implement the relaxation the embedding function $r(x,y,\tau)$ is discretized in $x$ and $y$, and eq.~(\ref{eq:r-tau}) is replaced by a set of ODEs for the values of $r$ at the lattice points, $r_{ij}(\tau)$.
We use $\tilde x = \tanh(x)$ to obtain a finite domain and a rectangular lattice with equidistant points. The derivatives are discretized using second-order finite differences and the boundary conditions are implemented such that they are compatible with the second-order accuracy of the finite differences.\footnote{%
For Neumann boundary conditions the lattice is extended by one row beyond the actual domain. The Neumann boundary conditions in the $y$ direction, (\ref{eq:Neumann-bc-y}), apply for regular points of $\partial\Sigma$, not at the locations of the 5-brane sources. This has to be taken into account in the discretization.}
The resulting set of ODEs is integrated numerically using Mathematica.

Asymptotically the evolution of the $r_{ij}(\tau)$ is expected to settle on an equilibrium configuration $r^\star_{ij}$, which is a discretized solution to the extremality condition for the minimal surface.
Letting the evolution (\ref{eq:r-tau}) run for a large time $\tau_{\rm max}\gg 1$ will yield an approximation to this equilibrium configuration. 
The quality of the final configuration $r_{ij}(\tau_{\rm max})$ can be assessed from the residuals
\begin{align}\label{eq:residuals}
	R_{ij}&=\left|L_\gamma^{-1}\frac{\delta L_\gamma}{\delta r(x,y,\tau)}\right|_{\tau=\tau_{\rm max}}~.
\end{align}
We typically use a lattice with $\mathcal O(100)$ nodes in the $\tilde x$ and $y$ directions,
though coarser lattices already capture the qualitative form of the surfaces well.
For the surfaces and data shown below the residuals have decreased to $\mathcal O(10^{-6})$ or less.
A limitation of this method is that it is unlikely to capture extremal surfaces for which the area functional does not take a local minimum (i.e.\ saddle points). However, the interest here is primarily in actual minimal surfaces.

Due to the symmetry of the D5/NS5 brane sources in the supergravity solutions (\ref{eq:h1h2-BCFT}) and (\ref{eq:h1h2-3d-grav}) under S-duality combined with $z\rightarrow i\pi/2 -z$, the Einstein-frame metric is invariant under $y\rightarrow \frac{\pi}{2}-y$. 
For the minimal surfaces discussed here the boundary conditions respect this symmetry, so that the surfaces themselves are symmetric.
The PDEs thus only have to be solved on the half of the strip $\Sigma$ with $0\leq y\leq\frac{\pi}{4}$, with a Neumann boundary condition at $y=\frac{\pi}{4}$ to enforce the symmetry.

\section{Islands with non-gravitating baths}\label{sec:islands}

In this section we discuss minimal surfaces, island contributions and the emergence of Page curves in the 10d solutions for non-gravitating baths, given in (\ref{eq:h1h2-BCFT}). 
The general structure of the supergravity solutions is illustrated in fig.~\ref{fig:AdS4-sol}, and we have D5 and NS5-brane sources, respectively, at $(x,y)=(0,0)$ and $(x,y)=(0,\pi/2)$.
The 8d minimal surfaces can be visualized as 2d surfaces in the 3d space spanned by the $x$ and $y$ coordinates parametrizing $\Sigma$ and the $AdS_4$ radial direction.
They are obtained using the relaxation method of sec.~\ref{sec:numerics}. 
We will start with a discussion of general features, before moving on to comparing the areas of island and HM surfaces.

\begin{figure}
	\includegraphics[width=0.4\linewidth]{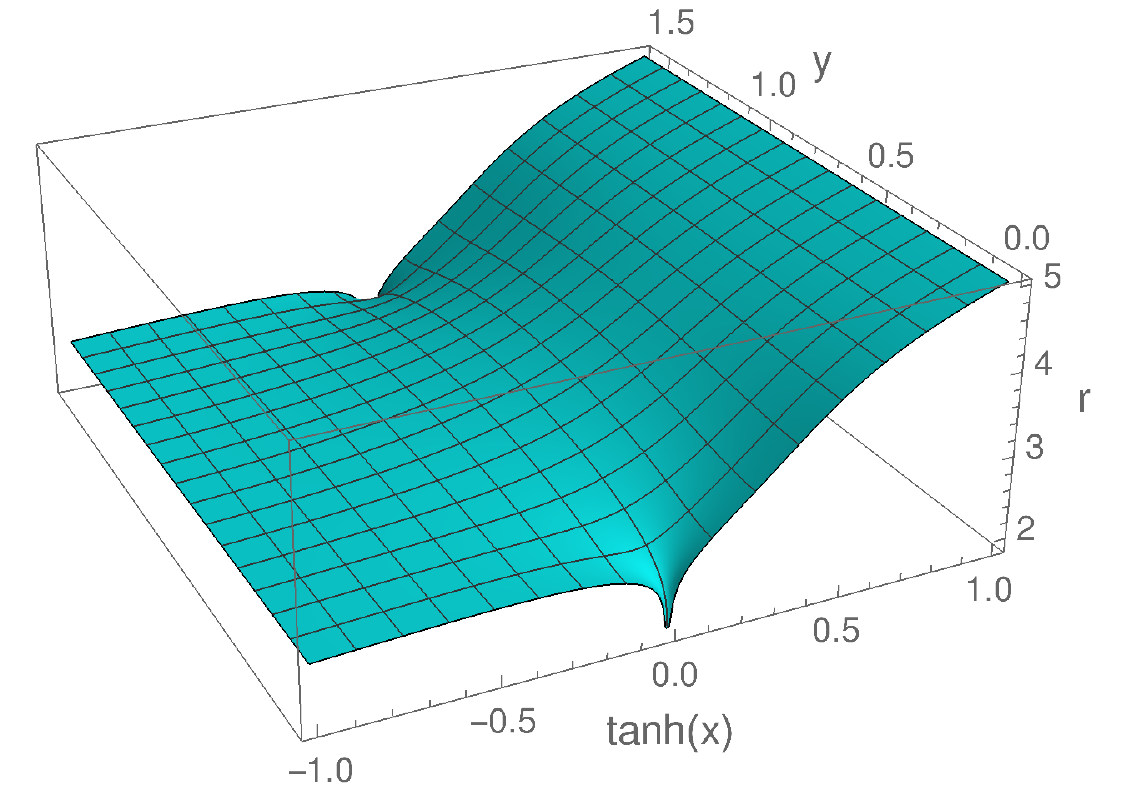}
	\hskip 10mm
	\includegraphics[width=0.4\linewidth]{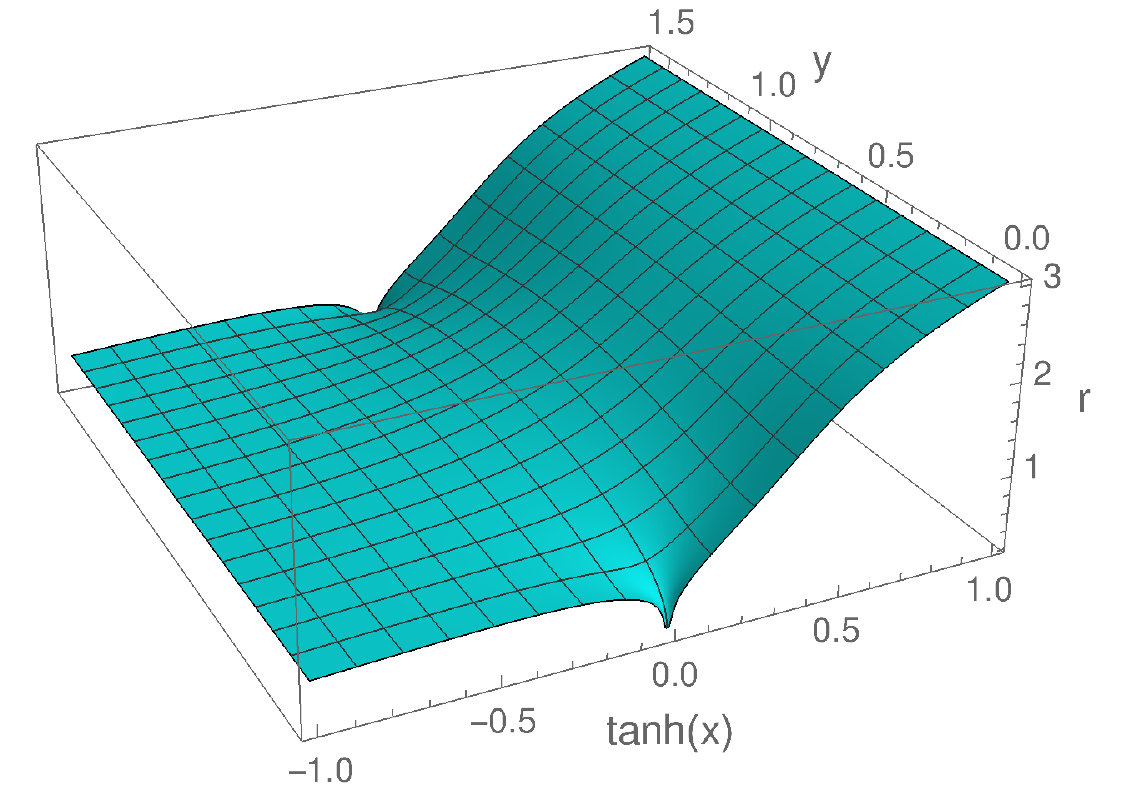}
	\\
	\includegraphics[width=0.4\linewidth]{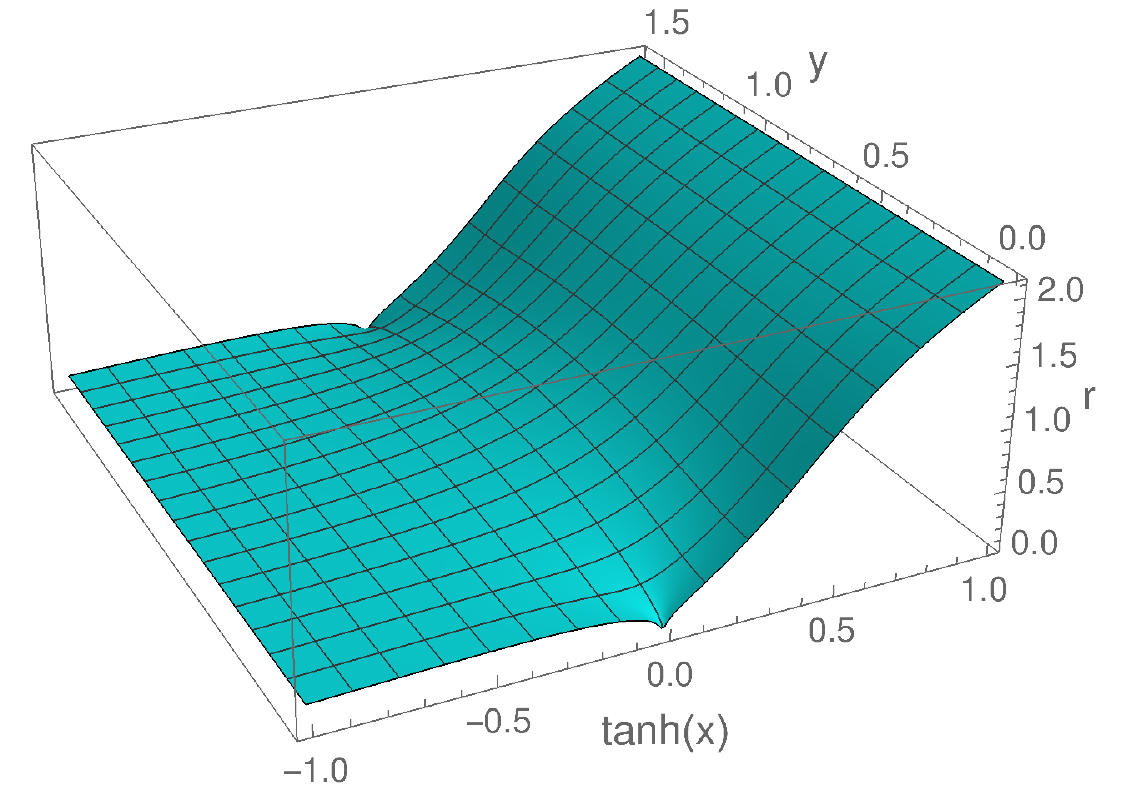}
	\hskip 10mm
	\includegraphics[width=0.4\linewidth]{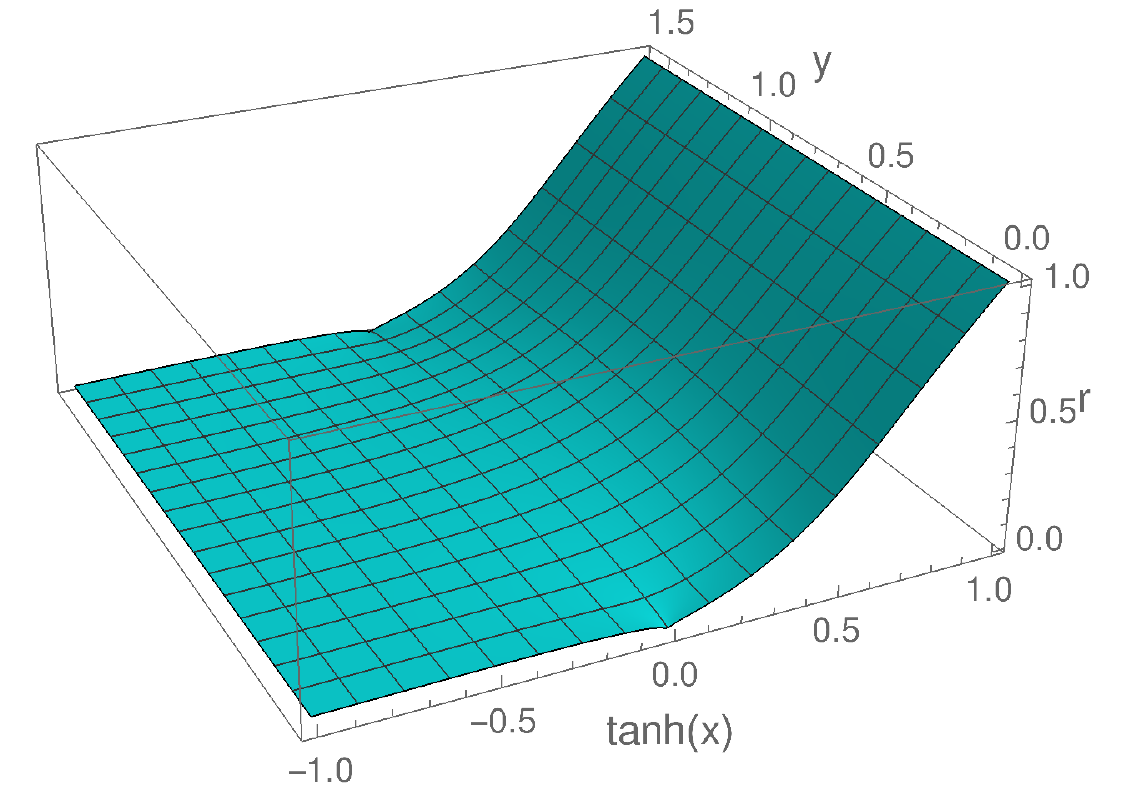}	
	\caption{
		Island surfaces from top left to bottom right anchored at $r_R\in\lbrace 5,3,2.1,1\rbrace$. The horizon is at $r_h=0$ and $N_5/K=2$. 
		The $AdS_5\times S^5$ region emerges at $\tanh x=1$, the 5-brane sources are at $\tanh x=0$ and $\tanh x=-1$ is a regular point in the internal space. For smaller $r_R$ (smaller radiation region) the surfaces stay closer to the horizon. Near the 5-brane sources the surfaces reach to the horizon for all $r_R$.
		\label{fig:islands}
	}
\end{figure}

\begin{figure}
	\includegraphics[width=0.43\linewidth]{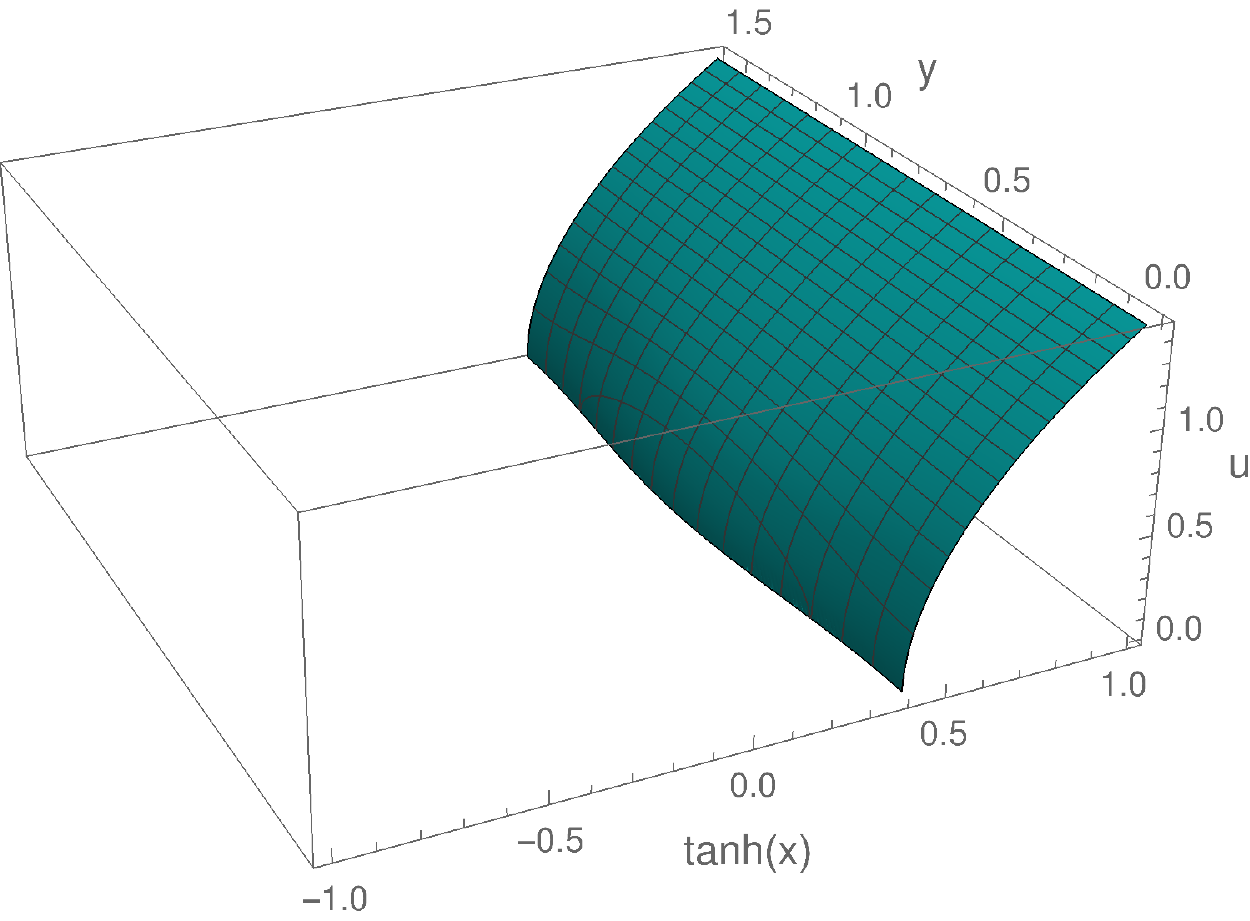}
	\hskip 10mm
	\includegraphics[width=0.43\linewidth]{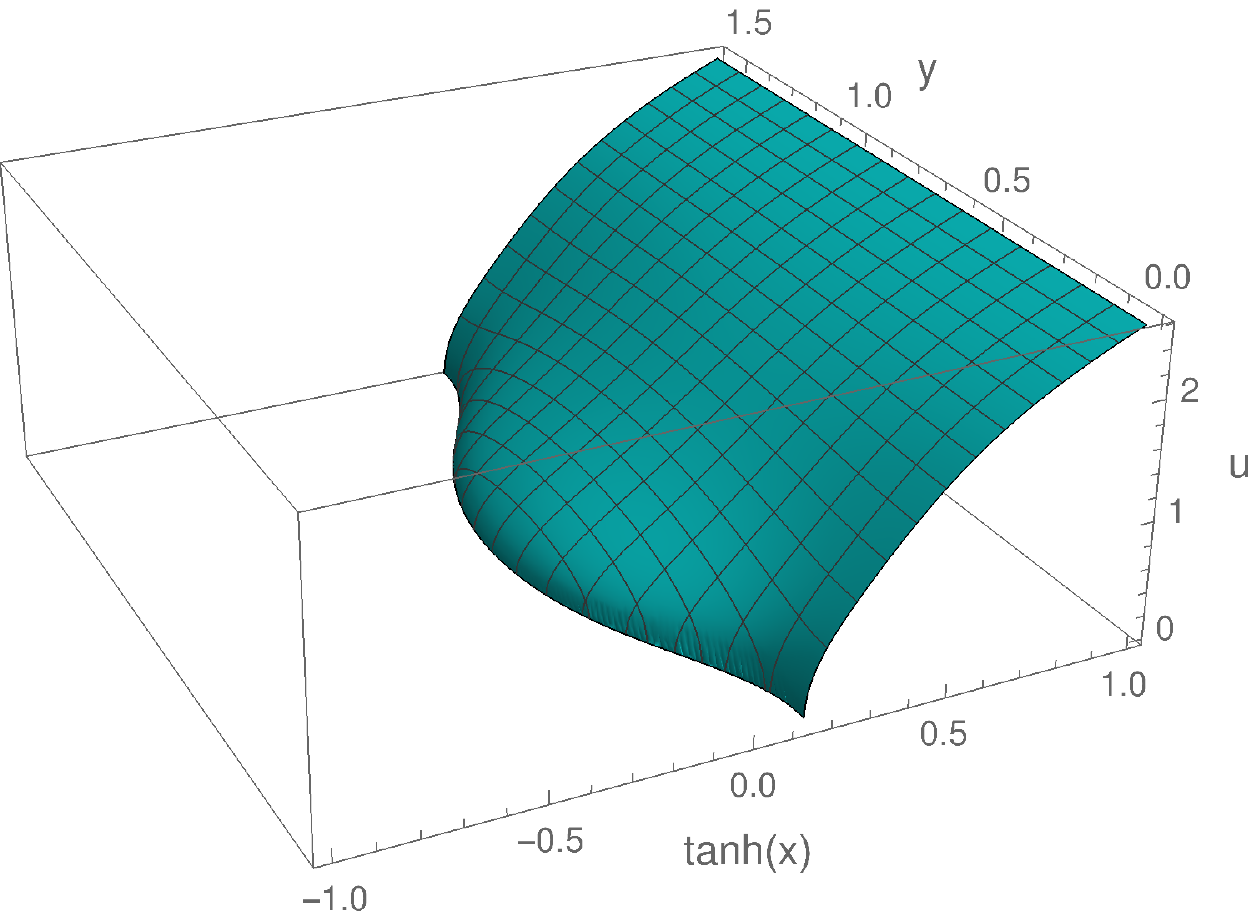}	
	\caption{
		HM surfaces at $t=0$ for $r_R=1$ (left) and $r_R=2.1$ (right), with $r_h=0$ and $N_5/K=2$. The plots show the tortoise radial coordinate $u$, in which the horizon is intersected orthogonally. 
		The further the HM surfaces are anchored from the horizon at $r_h=0$, the further they reach towards negative $x$.
		\label{fig:HM-surf}
	}
\end{figure}

\subsection{Island vs.\ HM surfaces}

A sample of island surfaces with varying anchor points $r_R=\lim_{x\rightarrow\infty}r(x,y)$ in the $AdS_5\times S^5$ region, for supergravity solutions (\ref{eq:h1h2-BCFT}), with temperature $r_h=0$ and $N_5/K=2$, is shown in fig.~\ref{fig:islands}. 
Simultaneous rescalings of $N_5$ and $K$ lead to an overall rescaling of the metric functions in (\ref{eq:10d-metric}), so the form of the minimal surfaces only depends on the ratio $N_5/K$. The ratio $N_5/K$ controls the ratio of the number of D3-branes suspended between the 5-branes and the number of semi-infinite D3-branes. For $N_5/K=2$ these numbers are equal (fig.~\ref{fig:brane-non-grav}).
For surfaces with large $r_R$, anchored far from the horizon, the impact of the 5-brane sources is clearly visible in fig.~\ref{fig:islands}, in line with the behavior discussed in sec.~\ref{sec:near-pole} (example surfaces at zero temperature are shown in fig.~\ref{fig:crit-T0-surfs}).
As the anchor point $r_R$ is decreased, moving towards the horizon,
the entire surface moves towards the horizon and the near-pole behavior becomes less pronounced.

For the surfaces in fig.~\ref{fig:islands} a discretization with $(200,100)$ points in $(\tanh x,y)$ was used, and the residuals (\ref{eq:residuals}) at $\tau=10^3$ are reduced to $\mathcal O(10^{-10})$. 
The quality of the solutions can also be investigated using the undiscretized extremality condition (\ref{eq:eom-fg}):
From a discretized solution one can construct a twice differentiable interpolating function $\tilde r(x,y)$. The interpolation should not necessarily be expected to capture the true solution accurately away from the lattice points, especially near the D5/NS5 sources where the true solution is not smooth. Evaluating (\ref{eq:eom-fg}) on the interpolation nevertheless only produces small errors near the poles, which decrease further with increased lattice resolution, suggesting that they are benign and not systematic.

Examples of $t=0$  HM surfaces for $N_5/K=2$  are shown in fig.~\ref{fig:HM-surf}.
For radiation regions that start far in the bath system, the surfaces are anchored close to the horizon at $\tanh x=1$, i.e.\ with $r_R$ close to $r_h$. These surfaces drop into the horizon along a curve $x_h(y)$ which is located well before reaching the D5/NS5 sources at $x=0$.\footnote{%
If the initial trial surface reaches beyond the 5-brane sources, the relaxation transitions it into the $x>0$ region.}
Upon moderately increasing $r_R$, the surfaces reach further towards smaller values of $x$.
The curve $x_h(y)$ starts to bulge out towards negative values in the interior of $\Sigma$, i.e.\ for $y\neq \lbrace 0,\pi/2\rbrace$, 
while the boundary values $x_h(0)$ and $x_h(\pi/2)$ remain at larger values and stay shy of reaching the 5-brane sources at $x=0$.
The behavior upon further increasing $r_R$ depends on $N_5/K$, and will be discussed below.

With the surfaces in hand we can compare the areas between island and $t=0$ HM surfaces anchored at the same $r_R$ and discuss the time evolution of the entropy.
The areas have the usual divergences associated with entanglement entropies in 4d.
Rather than isolating the divergences separately for island and HM surfaces, we directly compute the finite area difference between island and HM surfaces anchored at the same $r_R$. 
For numerical stability it is desirable to take the difference at the level of the integrands, at least in the region of large $x$.
Since the HM surface is obtained with a different parametrization, we transform the HM surface described by $x_{HM}(r,y)$ to a parametrization in terms of $r_{HM}(x,y)$, by inverting $x_{HM}(r,y)$ with respect to the first argument. 
The derivatives of $r_{\rm HM}$ can be expressed in terms of $x_{\rm HM}$,
\begin{align}
	\partial_x r_{HM}(x,y)&=\frac{1}{\partial_r x_{HM}(r,y)}\Big\vert_{r=r_{HM}(x,y)}~,
	&
	\partial_y r(x,y)&=-\frac{\partial_y x_{HM}(r,y)}{\partial_r x_{HM}(r,y)}\Big\vert_{r=r_{HM}(x,y)}~.
\end{align}
This is used to replace the derivatives in the area functional (\ref{eq:S}) before replacing $r_{HM}(x,y)$ itself by the inverse of $x_{\rm HM}$, to avoid taking derivatives of the inverted function and improve numerical stability.
The area integrands obtained this way are numerically smooth, and are used to compute the area differences with a cut-off $\tanh x\leq 1-\epsilon$.
The dependence on the cut-off is very mild, with percent level variation between $\epsilon=10^{-2}$ and $\epsilon=10^{-3}$,
and the latter is used for the plots.

\begin{figure}
	\begin{tikzpicture}
		\node at (0,0){\includegraphics[width=0.4\linewidth]{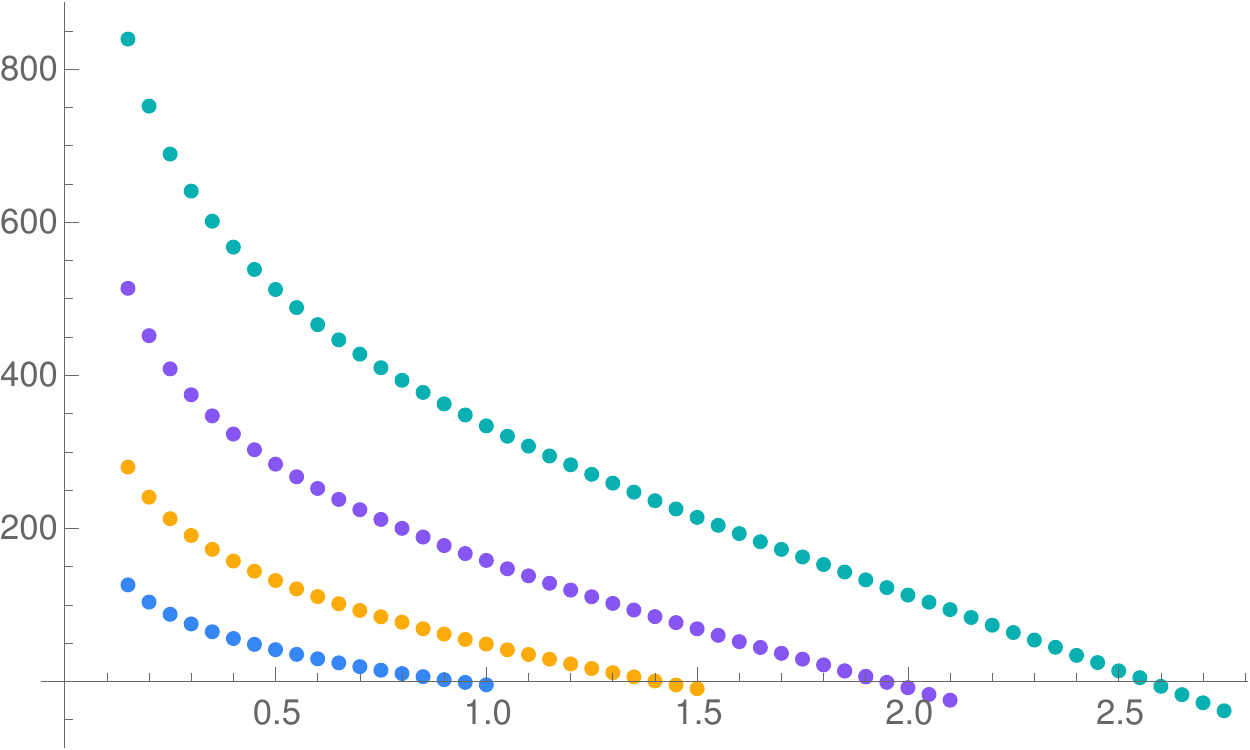}};
		\node at (3.6,-1.7) {\small $r_R$};
		\node at (-3.5,2) {\small $\Delta S$};
	\end{tikzpicture}

	\caption{Area difference $\Delta S=S_{\rm island}-S_{\rm HM}$ as function of the anchor point $r_R$ in the asymptotic $AdS_5\times S^5$ region. The defect is at $r=\infty$, the horizon at $r_h=0$.
		For $\Delta S>0$ the HM surface dominates at $t=0$.
		The radius of the $AdS_5\times S^5$ region is controlled by $N_5K$, the number of defect degrees of freedom by $N_5^2$.
		The color-coded dots are, from top to bottom, for $N_5/K\in\lbrace 1.2,1.6,2.0,2.4\rbrace$ with $K=1$.
		\label{fig:areadiff}}
\end{figure}

The results are shown in fig.~\ref{fig:areadiff}. They show that the island surface has larger area than the $t=0$ HM surface when $r_R$ is not too far from the horizon $r_h$. 
This leads to Page curves: The radiation region is identified far from the location where the gravity and bath systems meet ($r=\infty$), as the part of the $AdS_5\times S^5$ region at $x\rightarrow\infty$ with $AdS_4$ radial coordinate $r\leq r_R$, with $r_R$ close to $r_h$. 
The results in fig.~\ref{fig:areadiff} show that for these regions the area of the HM surface at $t=0$ is smaller than the area of the island surface. The area of the HM surface grows with time, but the entropy is bounded by the constant area of the island surface, leading to a Page curve.

The results in fig.~\ref{fig:areadiff} show that the area difference between the island and $t=0$ HM surfaces is larger for larger $N_5/K$.
One may compare this to expectations based on the Karch/Randall models: Larger $N_5/K$ corresponds to a BCFT with more 3d defect degrees of freedom relative to 4d bulk degrees of freedom, which in the 5d models amounts to larger tension of the ETW brane.
Larger tension bends the ETW brane towards the conformal boundary of $AdS_5$ in fig.~\ref{fig:KR} (i.e.\ smaller $\theta$; a tensionless brane has $\theta=\pi/2$).
From this 5d perspective one would expect the island surface to have larger area relative to the $t=0$ HM surface for smaller $\theta$, since the ETW brane is further from the bath. This is the 5d version of the area difference being larger for larger $N_5/K$ in 10d.

The curves in fig.~\ref{fig:areadiff} further show transition points $\hat r_R$ at which the areas of the island and HM surfaces are equal at $t=0$, suggesting constant entropies for $r_R>\hat r_R$. 
Near the end points of the curves, which for small $N_5/K$ are close to $\hat r_R$, the evolution of trial HM surfaces via (\ref{eq:r-tau}) changes: 
beyond values  $r_R^\star$ near the end points, the relaxation extends the trial surface all the way to $x=-\infty$ and ceases to settle on an equilibrium configuration.
If the HM surface becomes a shallow minimum or a saddle point, the relaxation could transition over it towards the island surface. 
One may also suspect that HM surfaces extending to negative $x$ also on the boundary of $\Sigma$ become relevant (those would reach to the horizon along a curve $x_h(y)$ and in a disconnected region around the 5-brane sources, and can not be parametrized globally by $x(u,y)$).
The value of $r_R^\star$ starts small for small $N_5/K$, increases to $r_R^\star\approx 2.1$ for $N_5/K=2$ 
(the surface on the right in fig.~\ref{fig:HM-surf} is close to $r_R^\star$), and appears to diverge towards $N_5/K\approx 4$. 
For larger $N_5/K$ HM surfaces can be found explicitly with no noticeable bound on $r_R$.
The limit $N_5\gg K$ corresponds to the number of 3d degrees of freedom being large compared to the number of 4d degrees of freedom. 
This corresponds in the 5d bottom-up models to an ETW brane close to the conformal boundary of $AdS_5$, which is the limit considered in \cite{Chen:2020uac,Chen:2020hmv}.
The separation between $r_R^\star$ and $\hat r_R$ appears to grow with $N_5/K$.
For radiation regions far in the bath (small $r_R$) we find Page curves.
For $r_R>\hat r_R$ fig.~\ref{fig:areadiff} suggests that the island surfaces lead to constant entropies,
though if new types of HM surfaces become relevant the entropy curve may remain non-trivial. 
In either case, the entropy is bounded by the area of the island surfaces, which we find explicitly for small and large $r_R$.

\subsection{Critical brane setups}

\begin{figure}
	\subfigure[][]{\label{fig:crit-T0-plot}
		\includegraphics[width=0.4\linewidth]{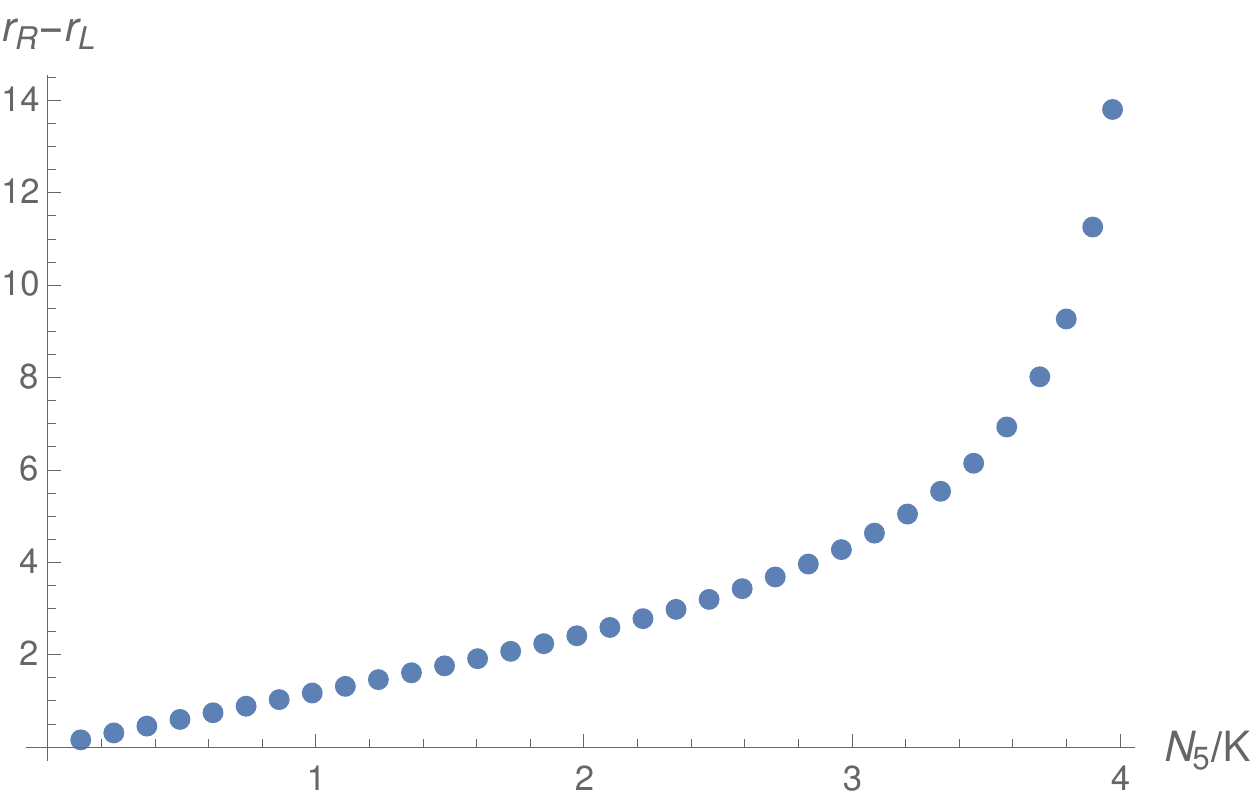}
	}
	\qquad
	\subfigure[][]{\label{fig:crit-T0-surfs}
		\includegraphics[width=0.4\linewidth]{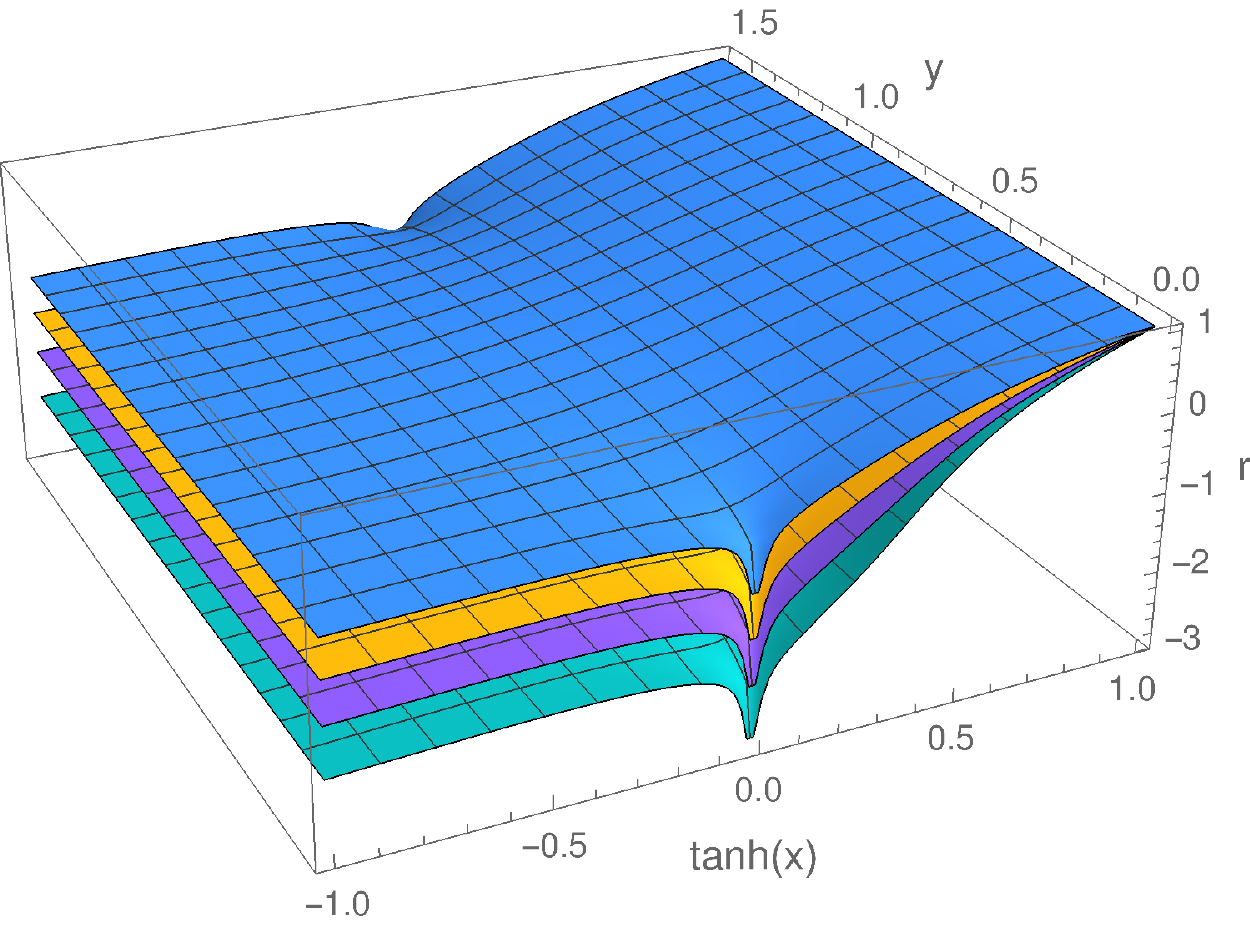}
	}
	\caption{Left: $r_R-r_L$, where $r_R= \lim_{x\rightarrow +\infty}r(x,y)$ is the  anchor of the minimal surface in the non-gravitating bath and $r_L= \lim_{x\rightarrow -\infty}r(x,y)$, as function of $N_5/K$ at zero temperature.
		Right: island surfaces, from top to bottom for $N_5/K\in\lbrace 1.2,1.6,2.0,2.4\rbrace$.
		At zero temperature $r_R-r_L$ is independent of $r_R$.
		\label{fig:crit-ang}}
\end{figure}

The ratio $N_5/K$ plays a prominent role also at zero temperature.
A sample of island minimal surfaces for different values of $N_5/K$ at zero temperature is shown in fig.~\ref{fig:crit-T0-surfs}.
For fixed anchor point in the asymptotic $AdS_5\times S^5$ region, the point where the surfaces close off at $x\rightarrow -\infty$ moves towards the Poincar\'e horizon as $N_5/K$ is increased. 
This is shown more quantitatively in fig.~\ref{fig:crit-T0-plot}, which shows the difference between $r_R=\lim_{x\rightarrow +\infty}r(x,y)$ and $r_L=\lim_{x\rightarrow -\infty} r(x,y)$ as function of $N_5/K$.
For small $N_5/K$ the difference $r_R-r_L$ grows linearly, but for larger $N_5/K$ it starts to grow rapidly.
The plot suggests the existence of a critical value, 
\begin{align}\label{eq:crit-nongrav-T0}
	\left(\frac{N_5}{K}\right)_{\rm crit}&\approx \ 4.0~,
\end{align}
at which $r_R-r_L$ diverges.
For the surfaces from which the data in fig.~\ref{fig:crit-T0-plot} is extracted the residuals (\ref{eq:residuals}) are reduced to at most $\mathcal O(10^{-7})$.
Increasing $N_5/K$ beyond the critical value appears to lead to irreducible residuals (\ref{eq:residuals}), which remain finite and keep driving the anchors $r_L$ and $r_R$ further apart with increasing runtime in $\tau$, rather than settling on an equilibrium configuration.
This is consistent with $r_R-r_L$ diverging when $N_5/K$ approaches (\ref{eq:crit-nongrav-T0}), and there being no island minimal surfaces beyond the critical value at zero temperature.

These results line up well with the observations in the Karch/Randall models: As noted before, the angle $\theta$ of the ETW brane in 5d is expected to be an effective description for the number of defect degrees of freedom relative to the number of 4d degrees of freedom, which in this particular example of a 10d solution is determined by the ratio $N_5/K$.
The discussion in \cite{Geng:2020fxl} found that, as the angle is decreased, the point where the island minimal surface with fixed anchor on the bath brane hits the ETW brane moves towards the infrared, and diverges towards the Poincar\'e horizon at a critical angle.
This is consistent with the behavior of the 10d solutions considered here if $1/\theta$ is identified with $N_5/K$.
It would be interesting to investigate more general 10d solutions, e.g.\ with multiple 5-brane sources, in which one may expect a more complicated phase structure.

\begin{figure}
	\includegraphics[width=0.4\linewidth]{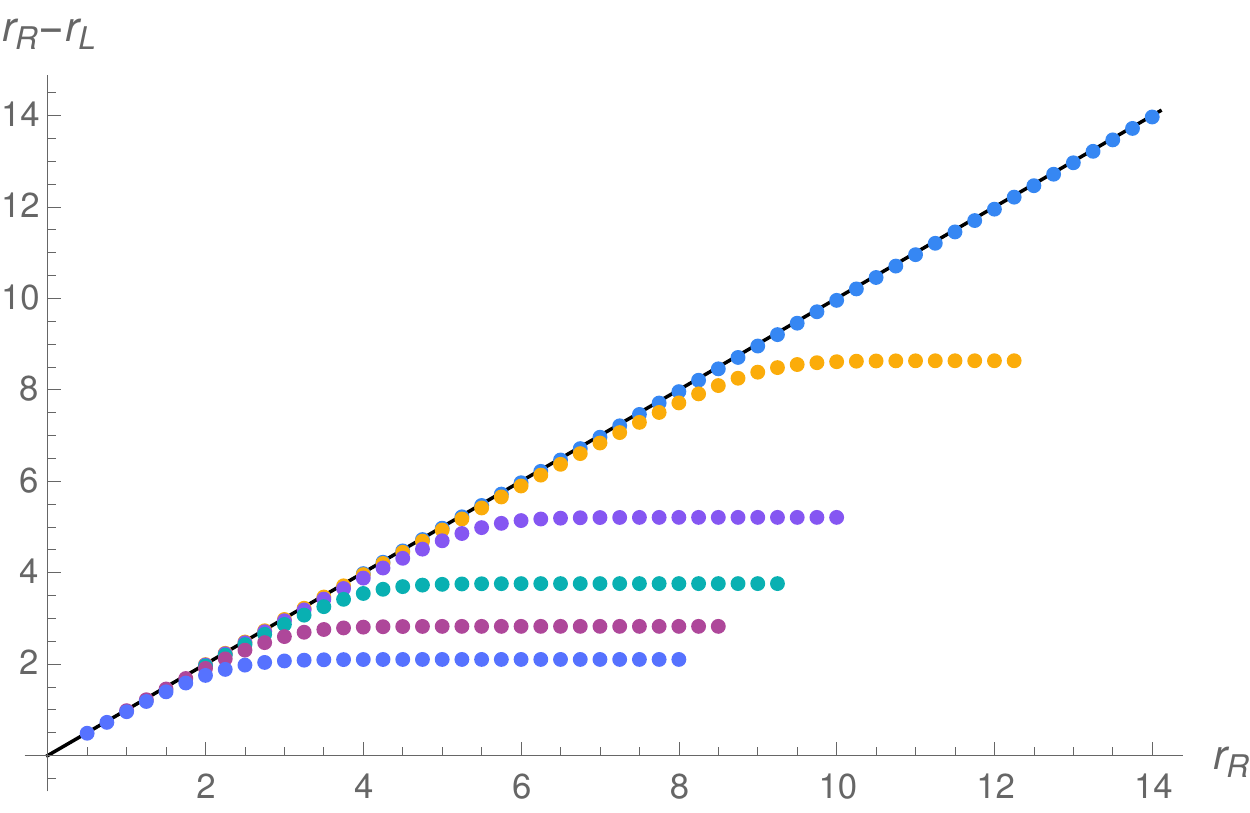}
	\caption{Difference $r_R-r_L$ between the end points at $x\rightarrow\pm\infty$ at finite temperature, with $r_h=0$, from bottom to top for $N_5/K\in\lbrace 2.25,2.75,3.25,3.75,4.25\rbrace$. The solid black line shows $r_L=0$.\label{fig:crit-ang-T}
	}	
\end{figure}

At finite temperature the runaway behavior of the cap-off point $r_L$ at the critical $N_5/K$ is regulated by the black hole horizon, and island surfaces can be found beyond the critical $N_5/K$.
The behavior can again be diagnosed by the difference $r_R-r_L$.
At zero temperature and below the critical $N_5/K$ this difference is finite and independent of $r_R$, with its value growing rapidly towards the critical $N_5/K$.
At finite temperature and below the critical value of $N_5/K$, the difference $r_R-r_L$ is not constant, but it approaches a constant for large $r_R$. 
This behavior can be seen in fig.~\ref{fig:crit-ang-T} as the curves that saturate towards a constant.
The constant is set by the zero temperature value of $r_R-r_L$.
As $N_5/K$ approaches the critical value, the point where the growth of $r_R-r_L$ saturates increases rapidly. 
The results are consistent with $r_R-r_L$ staying linear without bound for $N_5/K$ beyond the critical value.
The end point $r_L$ appears stuck below a critical value, similar to the behavior found in the 5d models in \cite{Geng:2020fxl}.

\section{Islands with gravitating baths}\label{sec:grav-bath}

We now turn to the gravitating bath solutions (\ref{eq:h1h2-3d-grav}), in which the $AdS_5\times S^5$ region at $x\rightarrow\infty$ is reduced to a regular point of the internal space (fig.~\ref{fig:AdS4-sol-grav}).
These solutions have massless 4d gravitons (the 4d Newton constant is related to the free energy of the dual 3d SCFTs which is proportional to $\int_\Sigma h_1 h_2 W$).
Without the $AdS_5\times S^5$ region there is no natural place to geometrically define radiation regions (compatible with diffeomorphism invariance)  at $x=\infty$, or to anchor minimal surfaces. 
Minimal surfaces stretching from $x=-\infty$ to $x=+\infty$ instead satisfy Neumann boundary conditions on both ends, as discussed in sec.~\ref{sec:bc}, and are found to settle onto the horizon. This leads to a flat entropy curve, in line with the arguments of \cite{Laddha:2020kvp} for gravitating baths.\footnote{Attempts to define notions of effective geometric entropies with dynamical gravity and discussions of their Page curves can be found in \cite{Krishnan:2020oun,Dong:2020uxp,Krishnan:2020fer}.}

As suggested in \cite{Laddha:2020kvp}, a Page curve may still arise for other quantities in situations with gravitating baths.
An alternative is to consider surfaces that divide the internal space, which may be expected to compute non-geometric EE's.
Though the general interpretation of such surfaces may not be entirely understood, one can view some of them in the current context as limiting cases of surfaces computing geometric EE's, as suggested in \cite{Geng:2020fxl} (an earlier example where geometric EE's turn non-geometric in the IR can be found in \cite{Balasubramanian:2017hgy}). The proposal of  \cite{Geng:2020fxl}  can be made precise in 10d: Consider brane configurations where D3-branes suspended between 5-branes are kept finite in extent, to realize $\mathcal N=4$ SYM on an interval. One may compute conventional geometric EE's on that interval. Though holographic duals for $\mathcal N=4$ SYM on an interval are not explicitly known, these geometric EE's would be represented by conventional Ryu/Takayanagi surfaces in the putative holographic duals. 
As IR fixed points one obtains 3d $T_\rho^\sigma[SU(N)]$ SCFTs, with holographic duals of the form discussed here.\footnote{
The setup can be seen as string theory realization of wedge holography in the sense of \cite{Akal:2020wfl}. The internal space in the 10d $AdS_4$ solution is the string theory uplift of the wedge region.
}
At the IR fixed point the geometric EE's on the interval become non-geometric EE's, and the Ryu/Takayanagi surfaces become minimal surfaces in the internal space.
We thus expect at least certain minimal surfaces separating regions in the internal space to compute non-geometric EE's.
In lower-dimensional examples such EE's are discussed in \cite{Geng:2021iyq}.

There are numerous ways to separate regions in the internal space in the 10d Type IIB solutions.
One may for example divide one of the $S^2$'s, which should be related to a split of the Hilbert space based on the $R$-symmetry  \cite{Karch:2014pma}. The surfaces which arise from geometric EE's as outlined above are expected to split the Riemann surface $\Sigma$ instead, where they can separate the 5-brane sources.
As shown in  \cite{Graham:2014iya}, minimal surfaces dividing the internal space end, when reaching the conformal boundary of the $AdS$ part, on an extremal sub-surface in the internal space. Boundary conditions can be imposed to fix the subleading behavior as the conformal boundary in the $AdS$ part is approached, instead of the leading behavior (e.g.\ for surfaces splitting the $S^5$ in $AdS_5\times S^5$ the slipping mode away from the equator).
In the solutions (\ref{eq:h1h2-3d-grav}) there is a natural candidate extremal surface in $\Sigma$: due to the reflection symmetry of the solution under $x\rightarrow -x$, the locus $x(y)=0$ is extremal in $\Sigma$ and can serve as an anchor point for 8d minimal surfaces wrapping the spatial part of $AdS_4$, both $S^2$'s and a curve in $\Sigma$ which depends on the $AdS_4$ radial coordinate $u$.

A symmetric HM surface which is anchored at $x(y)=0$ also in the thermofield double is given by $x(u,y)=0$. This entire surface is extremal thanks to the reflection symmetry in $x\rightarrow -x$.
More general surfaces may be obtained by specifying non-trivial subleading behavior in the $AdS_4$ radial coordinate as the $x=0$ locus is approached.\footnote{Admissible choices for the fall-off behavior near the boundary of $AdS_4$ can be determined by linearizing the extremality condition around the $x(u,y)=0$ surface and performing a mode expansion in the $y$ direction.}
We only impose that the surface be anchored at $x(y)=0$ for $u\rightarrow\infty$ and in the thermofield double, and then let the relaxation method settle on a surface. 
This procedure selects the $x(u,y)=0$ HM surface at $t=0$.
Since $x(y)=0$ is an extremal curve in $\Sigma$, finding the HM surfaces for $t\neq 0$ reduces to a problem within the $AdS_4$ part of the geometry, which is identical to the discussion in appendix A of \cite{Geng:2020fxl}.

\begin{figure}
	\includegraphics[width=0.3\linewidth]{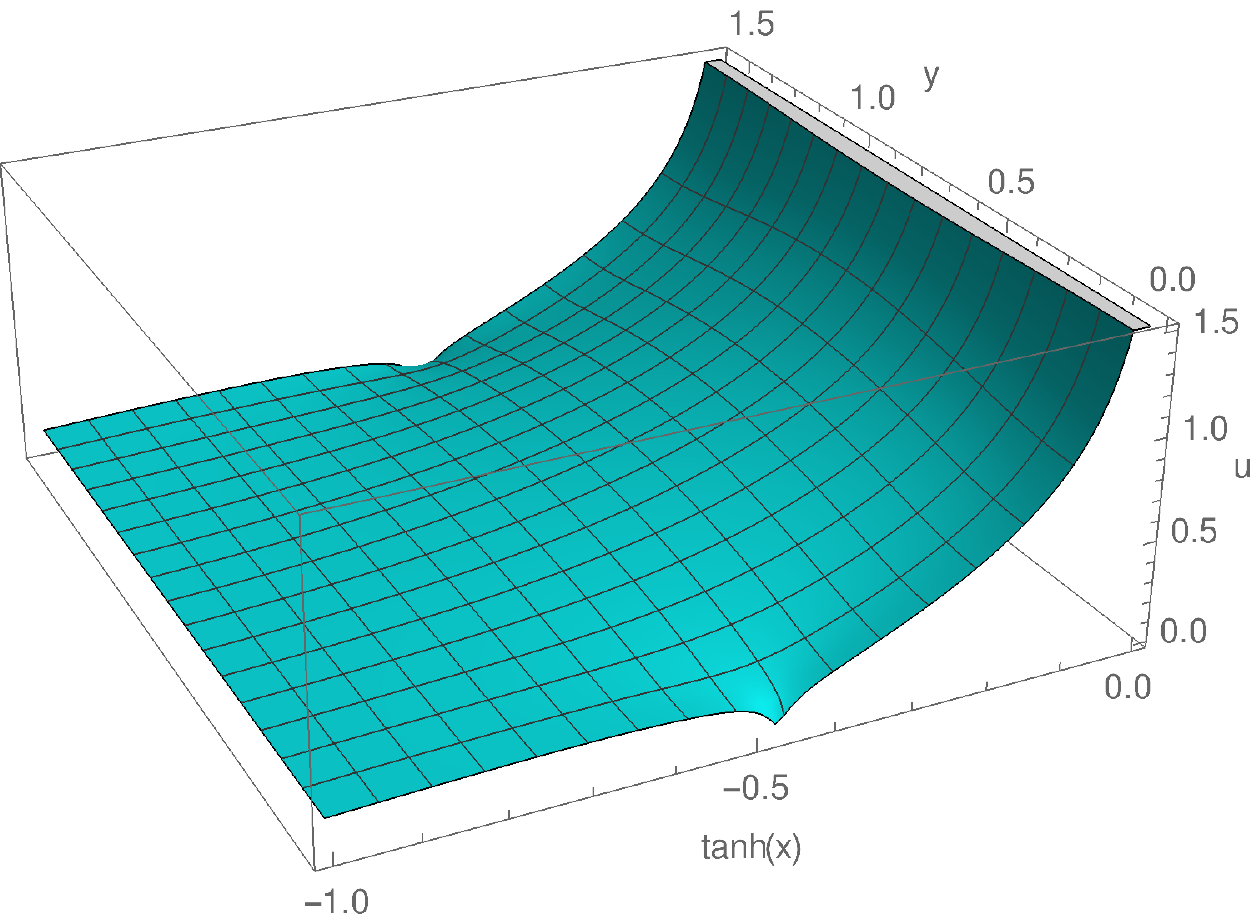}
	\hskip 5mm
	\includegraphics[width=0.3\linewidth]{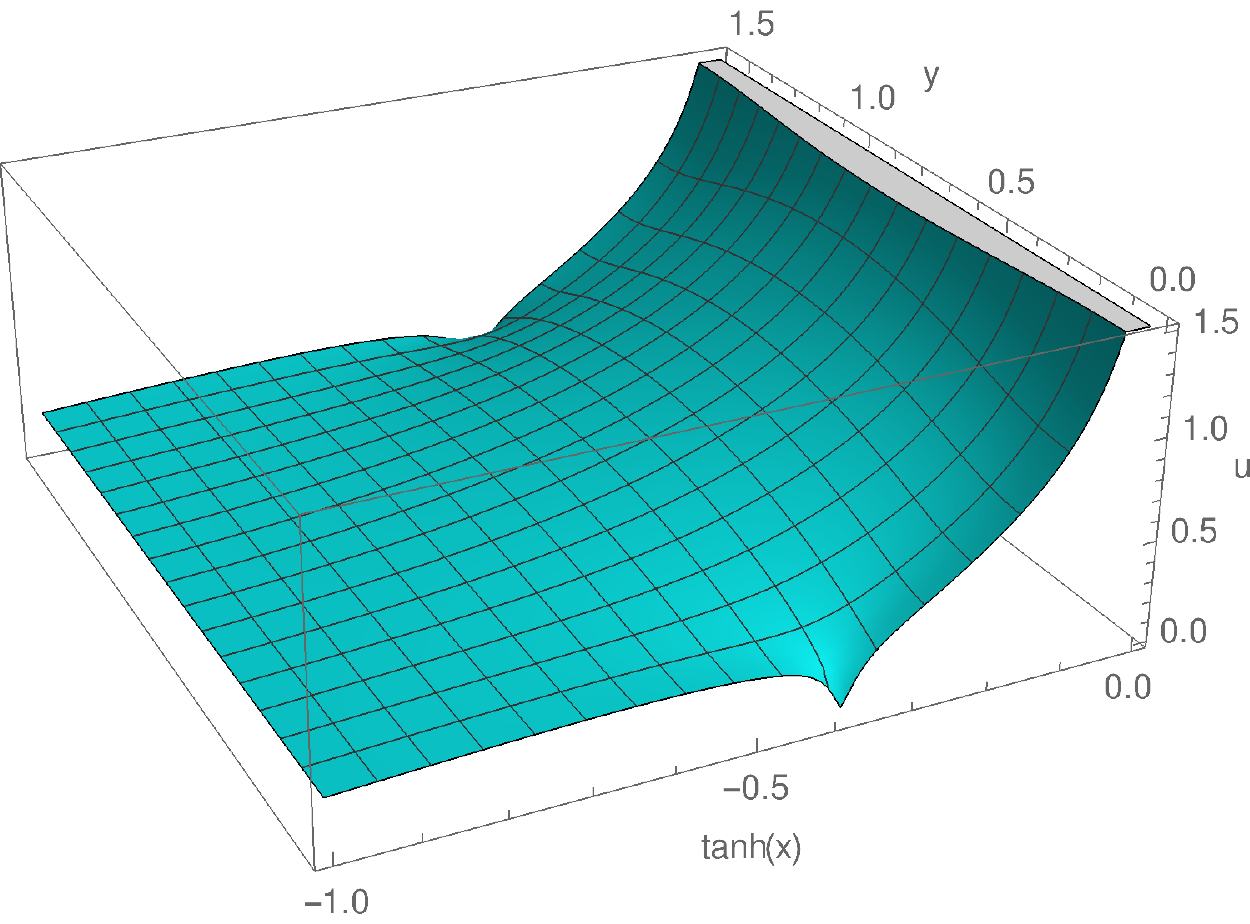}
	\hskip 5mm
	\includegraphics[width=0.3\linewidth]{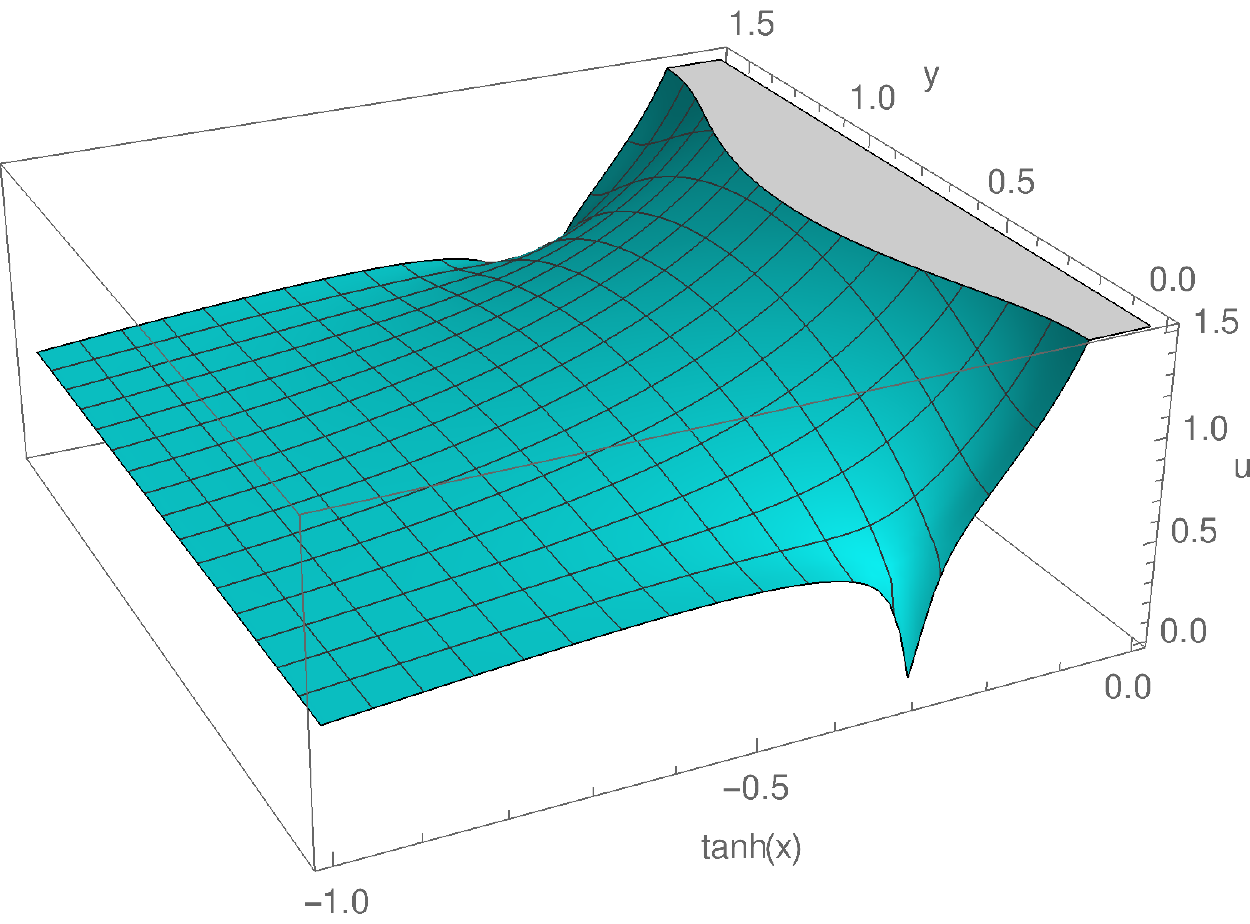}		
	\caption{Island surfaces in gravitating bath solutions, from left to right for $\delta\in\lbrace 0.5,0.4,0.3\rbrace$.
		The vertical axis shows the tortoise $AdS_4$ radial coordinate $u$. The surfaces are anchored at the conformal boundary of $AdS_4$ ($u\rightarrow\infty$) on the curve $x(y)=0$ in $\Sigma$.
		The plots only cover the $x\leq 0$ part of $\Sigma$. Near the 5-brane sources the surface caps off close to the horizon.  The cap-off point at $x=-\infty$ increases as $\delta$ decreases.
		\label{fig:LRcrit2}}
\end{figure}

For the island surfaces we impose that they are similarly anchored for $u\rightarrow\infty$ along the $x(y)=0$ curve.
They should reach one of the $x=\pm\infty$ regions with the Neumann boundary condition (\ref{eq:bc-x-minus}) for some value $u_L>0$, which is determined dynamically. Since the supergravity solution is invariant under $x\rightarrow -x$ the surfaces ending at $x=+\infty$ and $x=-\infty$ are symmetry-related, and we only construct the ones ending at $x=-\infty$ explicitly.

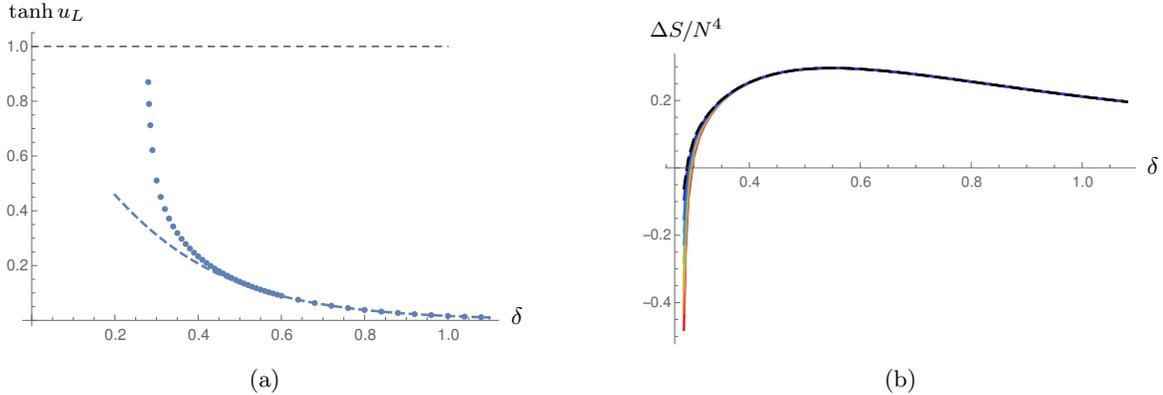
\begin{figure}
	\subfigure[][]{\label{fig:LRcrit1a}
		\begin{tikzpicture}
		\node at (0,0) {\includegraphics[width=0.4\linewidth]{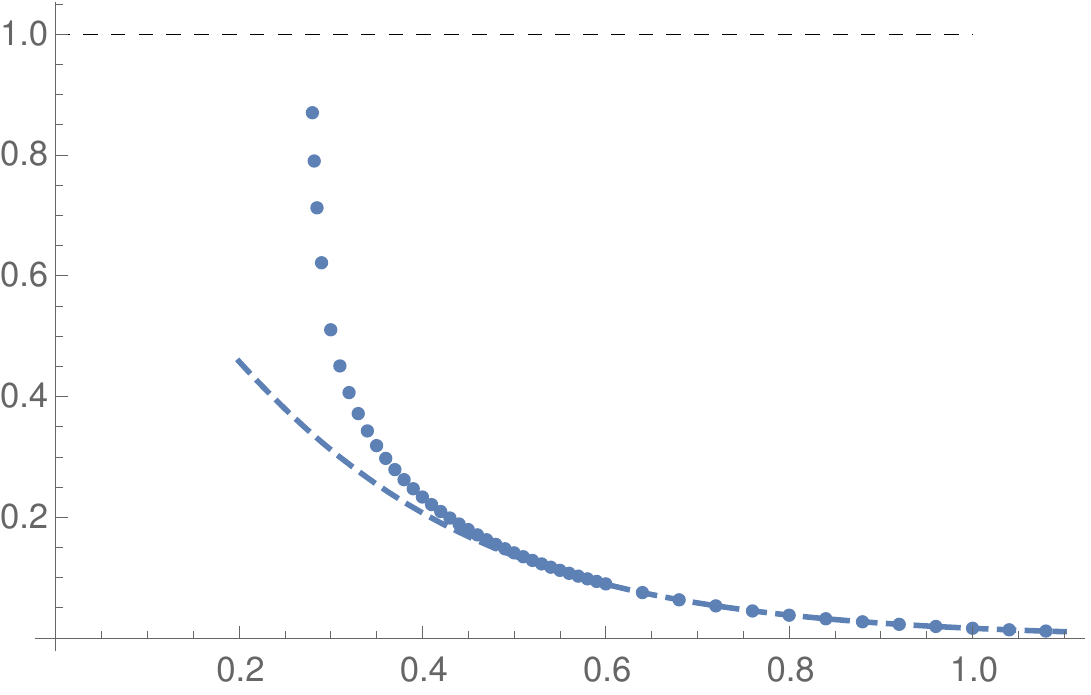}};
		\node at (-2.75,2.35) {\scriptsize $\tanh u_L$};
		\node at (3.5,-1.7) {\footnotesize $\delta$};
		\end{tikzpicture}
	}
	\hskip 10mm
	\subfigure[][]{\label{fig:LRcrit1b}
		\begin{tikzpicture}
		\node at (0,0) {\includegraphics[width=0.4\linewidth]{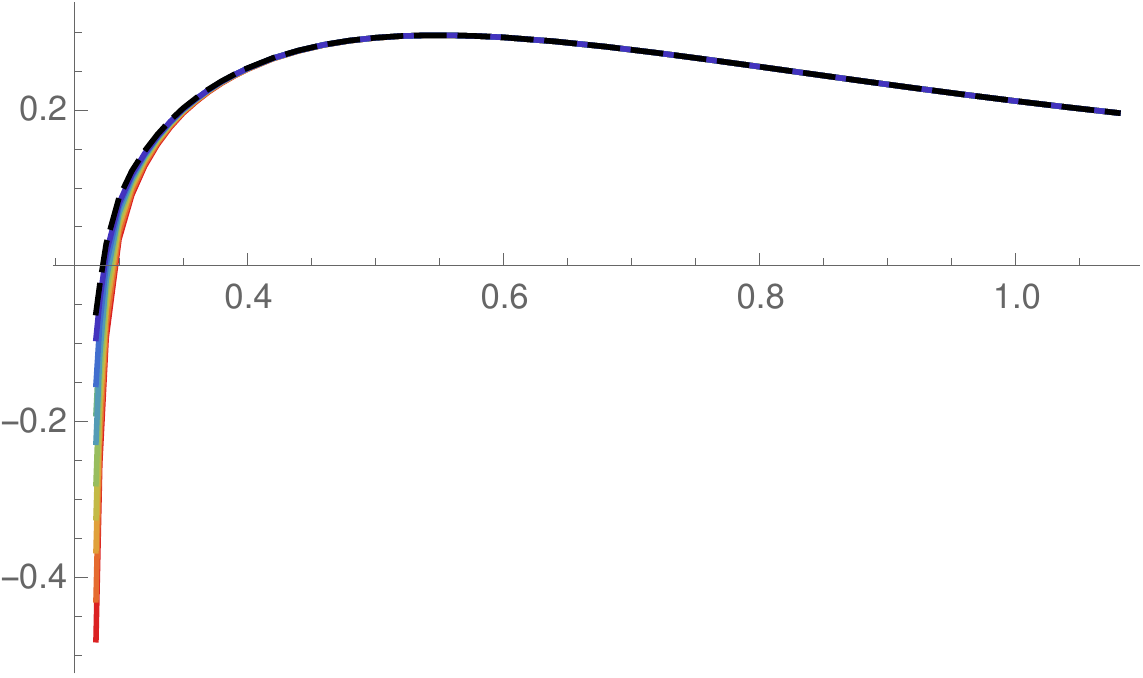}};
		\node at (-2.7,2.25) {\scriptsize $\Delta S/N^4$};
		\node at (3.5,0.45) {\footnotesize $\delta$};
		\end{tikzpicture}
	}
	\caption{Left: Cap-off point $u_L=\lim_{x\rightarrow -\infty} u(x,y)$ at $x=-\infty$ as function of the separation of brane sources $\delta$. 
		For large $\delta$, $u_L$ approaches the horizon at $u=0$ exponentially; the dashed line shows $u_L=1.17\exp(-4.28\,\delta)$. 
		At a finite $\delta_c$,  $u_L$ diverges towards the conformal boundary (at $\tanh u=1$ in the plot).
		Right: Area difference $\Delta S=S_{\rm island}-S_{\rm HM}$, as colored curves for different choices of cut-off on the $AdS_4$ radial coordinate. The dashed black curve shows an extrapolation to zero cut-off.
		\label{fig:LRcrit1}}
\end{figure}

A sample of island surfaces for different values of the 5-brane source locations $\delta$ on $\Sigma$ is shown in fig.~\ref{fig:LRcrit2} (the plots show only half the range of $x$).
For larger $\delta$ the surfaces more rapidly approach the horizon and then stay close to it.
This behavior is captured more quantitatively in fig.~\ref{fig:LRcrit1a}, which shows the end point at $x=-\infty$ in the tortoise coordinate $u$ as function of $\delta$.
The cap-off points $u_L$ show an exponential fall-off towards large $\delta$, which is shown as the fitted dashed line.
Towards small $\delta$ the cap-off points start to grow more rapidly.
The data is consistent with $u_L$ diverging towards the conformal boundary for a critical value
\begin{align}\label{eq:deltac}
	\delta_c&\approx 0.28~.
\end{align}
In line with this interpretation, the relaxation method does not settle on equilibrium minimal surfaces below $\delta_c$.
Instead, the trial surfaces keep approaching the conformal boundary of $AdS_4$ at generic points of $\Sigma$, while staying close to the horizon at the 5-brane sources (in line with behavior derived in sec.~\ref{sec:near-pole}). 
This will be discussed further below.

The area differences between the island and $t=0$ HM surfaces are computed similarly to the non-gravitating bath case. 
To implement the subtraction at the integrand level, the embedding for the island surface, $u_{\rm island}(x,y)$, has to be inverted with respect to the first argument to match the parametrization of the HM surface. The embeddings are not invertible on the entire domain, so the subtraction is implemented at the integrand level in a patch around $x=0$ and at the integral level for the remaining parts.
The resulting area differences are shown in fig.~\ref{fig:LRcrit1b}, as colored curves for different choices of cut-off on the $AdS_4$ radial coordinate. The cut-off on the radial coordinate is imposed in Fefferman-Graham gauge, $e^{-u}\geq\epsilon$ corresponding to $\tanh u\leq 1-2\epsilon^2$, with $\epsilon$ varied between $0.005$ and $0.05$.
The curves are indistinguishable for generic values of $\delta$. 
They only spread in a narrow region around $\delta_c$, where the island surfaces approach the conformal boundary of $AdS_4$ (though the cap-off point for the surfaces considered remains well below the cut-off) and residual cut-off dependence can be seen.
The residual cut-off dependence is smooth, and is fitted for each $\delta$ to obtain an extrapolation to zero cut-off.
The result is shown as dashed black curve.

The area differences in fig.~\ref{fig:LRcrit1} show that generically for large $\delta$ the HM surface at $t=0$ has smaller area than the island surface. The area of the HM surface grows in time, and when it equals that of the island surface the island surface becomes dominant, leading to a Page curve.
The curves in fig.~\ref{fig:LRcrit1} suggest a second distinguished value for $\delta$, 
a ``Page value" $\delta_P$ where $\Delta S$ at $t=0$ vanishes. 
The value of $\delta_P$ obtained from the numerical data,
\begin{align}
	\delta_P\approx 0.29\,,
\end{align}   
is close to but slightly larger than the critical $\delta_c$ in (\ref{eq:deltac}).
Since the difference between $\delta_c$ and $\delta_P$ is small and the island surfaces become numerically challenging for $\delta\approx \delta_c$, as evidenced in the spread of the curves in fig.~\ref{fig:LRcrit1},
the possibility remains that the true area difference may be non-negative for all $\delta>\delta_c$.
In the (possibly empty) range $\delta_c<\delta<\delta_P$, the island surface dominates already at $t=0$ and leads to a flat entropy curve.
Regardless of the relation between $\delta_c$ and $\delta_P$, for all $\delta>\delta_c$ the entropy growth indicated by the HM surface is limited by island surfaces whose area is constant.

\begin{figure}
	\subfigure[][]{\label{fig:DeltaSdC}
		\includegraphics[width=0.42\linewidth]{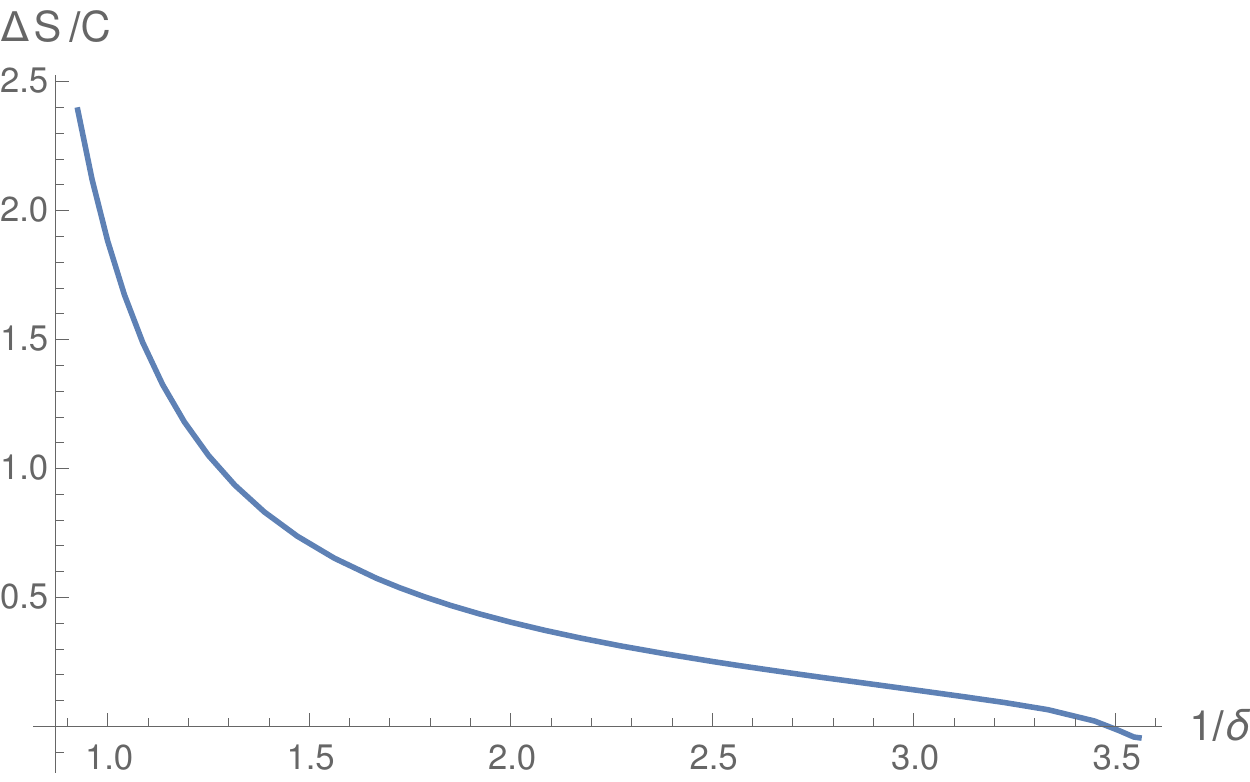}
	}\hskip 15mm
	\subfigure[][]{\label{fig:page}
		\includegraphics[width=0.4\linewidth]{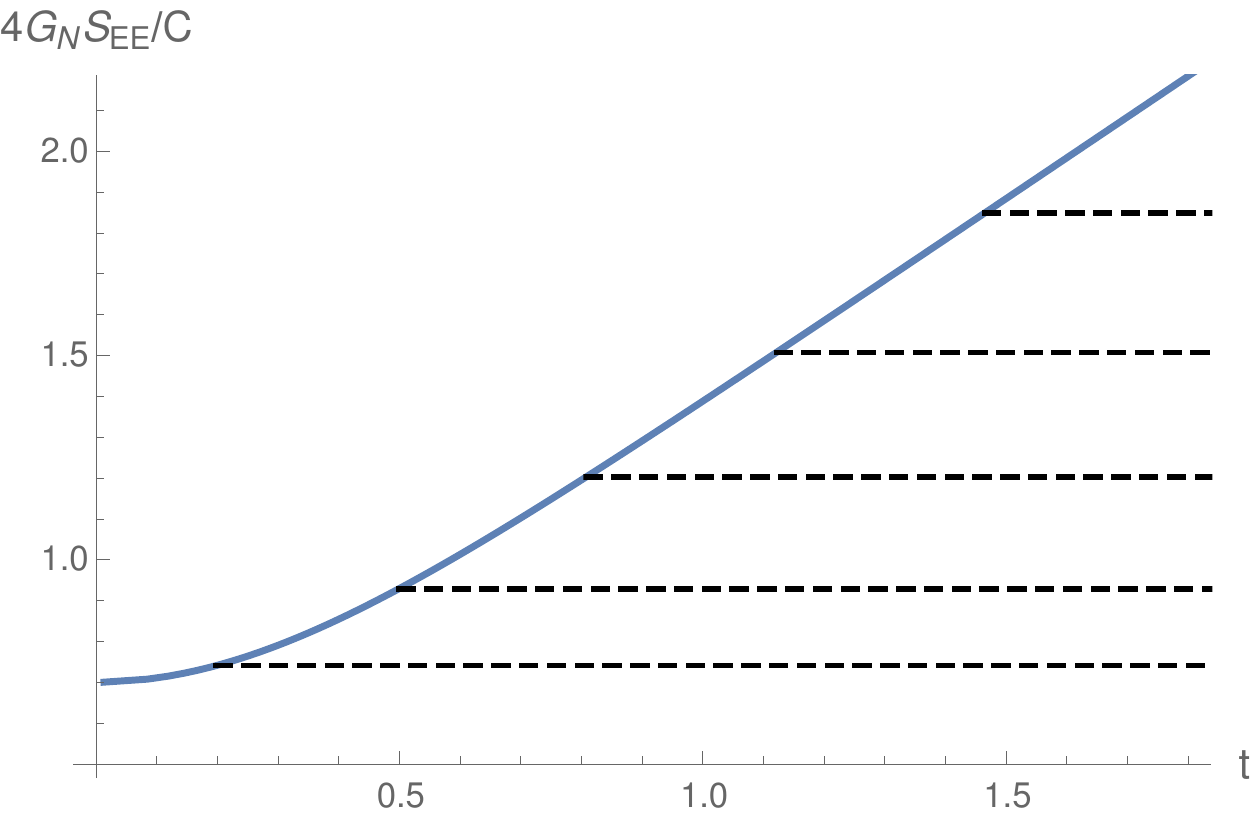}
	}
	\caption{Left: Area differences $\Delta S=S_{\rm island}-S_{\rm HM}$ normalized to the constant in (\ref{eq:C-def}). The plot shows the extrapolated curve of fig.~\ref{fig:LRcrit1b}.
		Right: Page curves. The solid line shows the time-dependent finite part of the area of the HM surface. The corresponding constant areas of island surfaces are shown as dashed lines, from bottom to top for $\delta\in\lbrace 0.29,0.32,0.4,0.5,0.6\rbrace$. The Page time increases monotonically with $\delta$.}
\end{figure}

Finding time-dependent HM surfaces reduces to a problem within the $AdS_4$ part of the geometry, since $x=0$ is an extremal curve in $\Sigma$. Up to an overall factor, the area as function of time can then be determined as in appendix A of \cite{Geng:2020fxl}, to which we refer for details on that part of the computation.
The overall factor arises from the parts of the internal space wrapped by the 8d minimal surfaces in the 10d solutions.
It can be determined by integrating the area functional in (\ref{eq:HM-area}) evaluated on the $x=0$ embedding over $y$.
This leads to the factor
\begin{align}\label{eq:C-def}
	C&=32\int_0^\pi dy\,\sqrt{\frac{1}{2}\left|h_1^3h_2^3W\right|}\,\Bigg\vert_{x=0}~.
\end{align}
It will be convenient to discuss the time-dependent entropy curves normalized to this factor, so that the (re)normalized area of the HM surface does not depend on the details of the 10d solution.
The area differences between island and HM surfaces at $t=0$ normalized to $C$ are shown in fig.~\ref{fig:DeltaSdC} as function of $1/\delta$.
The normalized area differences are monotonically increasing with $\delta$.
The time-dependent entropy curves, up to factors of $C$ and the 10d Newton constant, are shown in fig.~\ref{fig:page}. 
To obtain the curves a time-independent divergent part has been minimally subtracted, and a factor 2 has been included to account for the parts of the surfaces in the thermofield double.
Fig~\ref{fig:page} shows the transition from the HM surface to the island surfaces for various $\delta$.
The Page time, at which the transition occurs, increases monotonically with $\delta$: though the $t=0$ area differences in fig.~\ref{fig:LRcrit1b} are not monotonic, the Page time depends also on the growth rate of the HM surface, which decreases with $\delta$.
The Page time vanishes at $\delta_P$.

The 10d results are remarkably consistent with the phase structure found in 5d Karch/Randall models  if the inverse brane angle $\theta$ in 5d is seen as effective description for the brane stack separation $\delta$ on $\Sigma$:
the analysis of \cite{Geng:2020fxl} identified critical angles and Page angles, with a phase structure of minimal surfaces that, with the aforementioned identification, qualitatively matches the results found here (compare e.g.\ fig.~5 of \cite{Geng:2020fxl} to fig.~\ref{fig:DeltaSdC}). 
The symmetry of the 10d solutions (\ref{eq:h1h2-3d-grav}) under $x\rightarrow -x$ suggests that they give rise to Karch/Randall models with two equal ETW brane angles. 
More general Karch/Randall models with two unequal brane angles descend from 10d solutions with asymmetric distributions of 5-brane sources on $\Sigma$.

The range $\delta<\delta_c$, where no island minimal surfaces are found, corresponds to the regime above the critical ETW brane angle in 5d. The dominant contribution in 5d was identified as ``tiny islands", which arise as limiting surfaces that connect the defect to one of the ETW branes infinitesimally close to the conformal boundary.
In 10d the behavior of the island surfaces for  $\delta\rightarrow\delta_c$ and of the trial island surfaces below $\delta_c$, summarized around (\ref{eq:deltac}), both indicate that similar tiny island limiting surfaces arise for $\delta<\delta_c$.
The evolution of trial island surfaces for $\delta<\delta_c$ indicates that the 10d tiny islands approach the conformal boundary of $AdS_4$ almost everywhere on $\Sigma$, except for narrow throats around the 5-brane sources where they reach to the horizon.
In 5d the tiny islands were further motivated in \cite{Geng:2020fxl} through a deformation in which the ETW branes are separated and the tiny islands arise as limits of extremal surfaces computing geometric EE's.
This deformation has a clear analog in 10d, as keeping some D3 branes finite in extent to describe $\mathcal N=4$ SYM on an interval.
It would be interesting to study this deformation also in 10d, which would require as a first step the corresponding supergravity solutions.

\section{Outlook}\label{sec:outlook}

The results presented here  demonstrate in a UV-complete string theory setting the emergence of entanglement islands and Page curves for black holes in four-dimensional theories of gravity. 
The gravity theories certainly differ from the one we experience in nature. 
But they have dynamical gravitons, with a mass that can be controlled, and show versions of the information paradox whose resolution can be analyzed using concrete AdS/CFT dualities. We close with some thoughts on avenues for future exploration:

The discussions were based on representative Type IIB supergravity solutions that realize 5d Karch/Randall braneworlds with non-gravitating and gravitating baths in 10d.  
These solutions are members of a broad class of solutions corresponding to more general configurations of D3, D5 and NS5 branes. It would be interesting to study further examples.
The brane angles that play a crucial role in the phenomenology of the Karch/Randall models were given analogs in the representative 10d solutions, where the entanglement entropies exhibit a similar phase structure. One may suspect more complicated phase structures to emerge for more general 10d solutions.

It would be desirable to understand the time evolution of the entanglement entropies from the perspective of the dual QFTs.
The (critical) parameters in the supergravity solutions translate in a precise way to brane configurations and in turn to parameters in  $\mathcal N=4$ SYM on a half space and 3d $T_\rho^\sigma[SU(N)]$ SCFTs.
This should provide a concrete starting point for investigating the resolution of information paradoxes through entanglement islands in 4d using QFT methods.

A key holographic aspect appears to be a better understanding of minimal surfaces in the internal space and their associated field theory quantities. These are apparently  quantities which exhibit Page curve behavior with a gravitating bath, both in the 5d Karch/Randall models and in the string theory versions.
The 10d setups, with the full internal space present, should be a viable starting point for more detailed investigations of Page curve behavior in surfaces bisecting the internal space.
The surfaces studied in sec.~\ref{sec:grav-bath} are natural candidates for computing EEs associated with decompositions of the quiver diagram in the UV description of the dual 3d SCFTs.

\let\oldaddcontentsline\addcontentsline
\renewcommand{\addcontentsline}[3]{}
\begin{acknowledgments}
I am grateful to Andreas Karch,
Hao Geng, Carlos Perez-Pardavila, Suvrat Raju, Lisa Randall, Marcos Riojas, and
Sanjit Shashi for very interesting and useful discussions.	
This work is supported, in part, by the US Department of Energy under Grant No.~DE-SC0007859 	and by the Leinweber Center for Theoretical Physics.
\end{acknowledgments}
\let\addcontentsline\oldaddcontentsline

\bibliography{islands}
\end{document}